\documentclass[a4paper,prd,noshowpacs,noshowkeys,preprint]{revtex4}
\usepackage[latin1]{inputenc}
\usepackage{amsmath}
\usepackage{amssymb}
\usepackage{mathrsfs} 
\usepackage{hyperref}

\providecommand{\bp}{\mathbf{p}}

\providecommand{\bk}{\mathbf{k}}
\providecommand{\rk}{\mathrm{k}}
\providecommand{\bq}{\mathbf{q}}
\providecommand{\bQ}{\mathbf{Q}}
\providecommand{\bP}{\ms{\mathbf{P}}}
\providecommand{\bPF}{\ms{\mathbf{P}}_{\!F}}
\providecommand{\cM}{\mathcal{M}}
\providecommand{\meio}{\mss{\frac{\1}{\2}}}
\providecommand{\meios}{\mn{\frac{1}{2}}}
\providecommand{\TT}{{T_\Lambda}}
\providecommand{\TL}{{\mathcal{T}_\Lambda}}

\providecommand{\bx}{\mathbf{x}}

\providecommand{\lag}{\mathscr{L}}

\providecommand{\ponto}{{\cdot}}
\providecommand{\bnu}{{\bar{\nu}}}
\providecommand{\nue}{{\nu_e}}
\providecommand{\bnumu}{{\bar{\nu}_\mu}}
\providecommand{\zero}{{\mt{(0)}}}
\providecommand{\cS}{{\mathcal{S}}}

\providecommand{\eq}[1]{\begin{equation} #1 \end{equation}}
\providecommand{\eqarr}[1]{\begin{eqnarray} #1 \end{eqnarray}}
\providecommand{\ms}[1]{\mbox{\small $#1$}}
\providecommand{\mt}[1]{\mbox{\tiny $#1$}}
\providecommand{\mss}[1]{\mbox{\scriptsize $#1$}}
\providecommand{\mn}[1]{\mbox{\normalsize $#1$}}
\providecommand{\mfn}[1]{\mbox{\footnotesize $#1$}}
\providecommand{\ml}[1]{\mbox{\large $#1$}}
\providecommand{\mL}[1]{\mbox{\Large $#1$}}
\providecommand{\bs}[1]{\boldsymbol{#1}}
\providecommand{\bra}[1]{\langle #1 \vert}
\providecommand{\ket}[1]{\vert #1 \rangle}
\providecommand{\braket}[2]{\langle #1\vert #2 \rangle}
\def\1{\mbox{\tiny $1$}}
\def\2{\mbox{\tiny $2$}}
\providecommand{\hs}[1]{\hspace{#1}}
\providecommand{\re}{\mathrm{Re}}
\begin{document}
\title{
Intrinsic flavor violation in neutrinos produced through decays
}
\author{C.~C.~Nishi}
\email{celso.nishi@ufabc.edu.br}
\affiliation{
Universidade Federal do ABC - UFABC\\
Rua Santa Adélia, 166, 09.210-170, Santo André, SP, Brazil
}

\begin{abstract}
We show there is a non-null probability to produce neutrinos of the ``wrong'' type
in general decays, within the SM augmented by the known
three massive neutrinos and nontrivial mixing.
Such effect is equivalent to an intrinsic flavor violation 
(lepton flavor violation) at creation without requiring neutrino propagation.
The exact amount of flavor violation depends on the neutrino flavor state
to be detected. For realistic conditions, the violation is tiny but much larger than
other indirect lepton flavor violation processes not involving neutrinos as final
states. 
For neutrinos produced in a continuous spectrum, we show the effect is
larger, in relative terms, for neutrinos produced in the low-energy portion of the
spectrum.
For muon decay, if the usual neutrino flavor state is assumed a flavor violation as
large as 1\% is possible in the channel $\mu^+\to\bnu_\tau e^+ \nu_e$.
We also discuss the relation between flavor violation and flavor indefiniteness
for neutrino flavor states.

\end{abstract}
\maketitle
\setlength{\baselineskip}{.92\baselineskip}
\section{Introduction}
\label{sec:intro}

Family lepton numbers or lepton flavors ($L_e,L_\mu,L_\tau$) are conserved
in the SM to a great extent. The clearest sign of nonconservation appears in the
neutrino sector where neutrinos where observed to undergo flavor conversion as they
propagate long distances.
We have inferred from such phenomenon that neutrinos are massive and mix themselves 
through large mixing angles.
As such, lepton flavors are not exactly conserved quantum numbers within the SM
augmented by massive neutrinos\,\cite{bilenky,casasibarra,LFV,endnote0}.
Nevertheless, the amount of indirect lepton flavor violation induced by neutrino
masses, such as in $\mu\to e\gamma$, is extremely small and
unobservable\,\cite{casasibarra} in comparison to the direct effect of neutrino
oscillations.

Structurally, the existence of replicated families of quarks and leptons by itself is
an unexplained feature of the SM.
As far as electromagnetic and strong interactions are concerned each particle with
the same electric charge behaves in the same way. Very different mass scales (and
then the interaction with the Higgs), however, distinguish the three known families.
The presence of weak interactions introduces the quark mixing responsible for flavor
violating processes. Quark flavor, however, are identified through the
different quark masses, despite the impossibility to observe them freely.
For massless neutrinos, the definition of neutrino flavor is intimately connected
to the definition of lepton flavor, which can be defined exactly as a conserved
quantum number, at least at the classical level.
With the presence of tiny neutrino masses, lepton flavors can still be
considered as approximately conserved quantities and neutrino flavors correspond to
the superpositions of mass eigenstates (fields) that 
carry the lepton flavor in the massless neutrino limit.
In that respect, the definition of neutrino flavor is unique within the SM in the
sense that it is defined as a superposition of mass eigenstates. Such structure is
responsible for neutrino flavor oscillations.

We intend to consider here a slightly different type of neutrino flavor violation,
\textit{i.e.}, an intrinsic flavor violation that could be present without
propagation.
As neutrino flavors are defined, in some approximate way, as certain
superpositions of massive neutrino states, some level of flavor indefiniteness is
expected at least at the order of $(\Delta
m/E_\nu)^2$\,\cite{giunti:torino04,review}.
If a strict definition is considered we can explicitly calculate the amount of
flavor violation that arises. Such task was undertaken previously for neutrinos
created in pion decay\,\cite{intrinsic}.	
In that work we have shown that the probability for the flavor violating
channel $\pi\to\mu\nu_\beta$, $\beta\neq\mu$, is given by
\eq{
\label{Pmunue:final}
\mathcal{P}\ms{(\pi\to\mu\nu_\beta)}\approx
\frac{1}{2}\sin^2\!2\theta
\Big(\frac{\Delta m^2}{2E_\nu\Gamma}\Big)^2
\,.
}
The largest effect was found for $\beta=\tau$ for which the probability was of the
order of $10^{-6}$. Although very small, such effect is much larger than the
branching ratio of indirect flavor violating processes such as $\mu\to e\gamma$.

In this work, we want to generalize such treatment to calculate the amount of
flavor violation in general decays, in special, in muon decay.
The generalization occurs in the sense of considering decays involving 
three or more decay products, since pion decay is a two-body decay that was 
kinematically easier to treat.
This generalization is important in two respects. Firstly, pion and muon decays
are the major sources of terrestrial accelerator neutrino experiments and
atmospheric neutrinos. Second, it is important to check how the neutrino flavor
violation effect changes with a continuous emission spectrum for the neutrino.
For that reason, the emission spectrum for the production of the wrong neutrino
flavor is also calculated.
It is known that the emission spectrum for $\nu_e$ in muon decay agree
experimentally\,\cite{mudecay:exp} with the SM prediction\,\cite{greub:plb93} and
the effects of neutrino masses introduces negligible distortions.

Ultimately, this work intends to clarify the validity of the usual definition of
neutrino flavor and its dependence with properties such as localization aspects
that is known to play a crucial role in neutrino oscillations\,\cite{kayser:81}.
Recent discussions on the oscillation of Mossbauer
neutrinos\,\cite{lindner:mossbauer} or entangled neutrinos\,\cite{glashow:no} shows
localization is an important aspect and should be carefully analyzed.
We are concerned, however, with a related but more fundamental
issue, \textit{i.e.}, the definition of neutrino flavor and its universality.

The outline of the article is as follows: in Sec.\,\ref{sec:decay} we derive the
general formula for neutrino flavor violation in decays. In
Sec.\,\ref{sec:mudecay}, we calculate the neutrino flavor violation in muon decay.
In Sec.\,\ref{sec:DeltaE=0} we define equal-energy neutrino flavor states and 
calculate the flavor violation probability for them.
Discussions and conclusions are presented in Sec.\,\ref{sec:discussion}.
The appendices contain auxiliary material that were chosen to be separated from
the main text.

\section{Neutrino flavor violation in decay}
\label{sec:decay}

Within the SM, ordinary neutrinos can be produced from two elementary processes: (a)
charged current processes (involving $W^\pm$) and neutral current processes
(involving $Z^0$). For the latter case, one neutrino and one antineutrino (two
neutrinos) are produced in accordance to lepton number conservation in
the SM but still neutrino oscillations might be possible\,\cite{smirnov:Z0}.
We will be interested here only in neutrinos produced through decays induced
by charged currents.

We will assume through this and the next section that neutrino and antineutrino
flavor states are approximately well described by the
superpositions\,\cite{giunti:torino04,intrinsic}
\eqarr{
\ket{\nu_\beta(\bk)}&\equiv&
U^*_{\beta j} \ket{\nu_j(\bk)}
\,,\cr
\label{nubeta}
\ket{\bnu_\beta(\bk)}&\equiv&
U_{\beta j} \ket{\bnu_j(\bk)}
\,.
}
where $\ket{\nu_j(\bk)}$ and $\ket{\bnu_j(\bk)}$ are well defined asymptotic
states with definite masses\,\cite{endnote2}, normalized as
$\braket{\bk}{\bk'}=\delta^3(\bk-\bk')$. The latter normalization will be used for
all momentum defined states henceforth.
The orthogonality among the mass eigenstates $\ket{\nu_j(\bk)}$ implies
\eq{
\label{nu:orto}
\braket{\nu_\alpha(\bk)}{\nu_\beta(\bk')}
=\delta_{\alpha\beta}\delta^3(\bk-\bk')\,.
}

Furthermore, we are interested in assessing the probability of lepton flavor
violation for neutrino flavor states \eqref{nubeta} produced in decays. We can
distinguish two types:
\eqarr{
\label{typeA}
\text{(A)} && l_\alpha\rightarrow \nu_\beta+X\,, \\
\label{typeB}
\text{(B)} && I\rightarrow l^+_\alpha+\nu_\beta+X\,,
}
where $X$ is one or more particles without net lepton number.
There is lepton flavor nonconservation, in the sense of an approximate
family lepton number, if $\beta\neq\alpha$.
Types A and B of flavor nonconservation are classified according to the flavor
nonconservation between (A) an initial and a final particle or (B) between two final
particles.
Pion decay is an example of type B decay without the accompanying product $X$. The
neutrino flavor violation in that context was calculated
previously\,\cite{intrinsic}.
Muon decay is an example where both types of violation may occur and it will be
treated in Sec.\,\ref{sec:mudecay}.
Usually, however, only one type of violation will be possible.
The CPT conjugate processes of \eqref{typeA} and \eqref{typeB} can be equally
considered.

Because of the definitions in Eq.\,\eqref{nubeta}, the amplitude for these decay
channels in Eqs.\,\eqref{typeA} and \eqref{typeB}, with fixed $\alpha$, should be
regarded as a coherent sum of the amplitudes of the 3 channels involving the neutrino
mass eigenstates $\nu_i$, \textit{i.e.},
(A) $l_\alpha\rightarrow \nu_i+X$ or (B) $I\rightarrow l^+_\alpha+\nu_i+X$,
$i=1,2,3$.
The weight of each channel is dictated by the SM weak interactions, considering
non-trivial mixing and non-degenerate masses for the three families of neutrinos
$\nu_i$.

Once we have the time evolution of the decaying state $\ket{l_\alpha(t)}$ or
$\ket{I(t)}$, the flavor violation probability can be calculated by
\eqarr{
\label{FvioA:1}
\mathcal{P}_{l_\alpha\to\nu_\beta}(t)=
\int\! [d^3\bP_F]\, 
|\braket{X,\nu_\beta}{l_\alpha(t)}|^2
&=&
\int\! [d^3\bP_F]\, 
|\sum_jU_{\beta j}\braket{X,\nu_j}{l_\alpha(t)}|^2
\,,
\\
\label{FvioB:1}
\mathcal{P}_{\to\bar{l}_\alpha\nu_\beta}(t)=
\int\! [d^3\bP_F]\, 
|\braket{X,\bar{l}_\alpha,\nu_\beta}{I(t)}|^2
&=&
\int\! [d^3\bP_F]\, 
|\sum_jU_{\beta j}\braket{X,\bar{l}_\alpha,\nu_j}{I(t)}|^2
}
where $\int[d^3\bP_F]$ denotes the integration over all final momenta.
The sum over final spin states is implicit.
We can associate a wave packet $\psi_I$ to the parent particle, encoding
informations on momentum distribution and spatial localization, by setting as the
initial states
\eqarr{
\label{la:psi}
\ket{l_\alpha}&=&\int d^3\bp\, \psi_I(\bp)\ket{l_\alpha(\bp)}\,,
\\
\label{I:psi}
\ket{I}&=&\int d^3\bp\,\psi_I(\bp)\ket{I(\bp)}\,.
}
We can assume that for a time $t\gg 1/\Gamma$, where $\Gamma$ is the decay
width of the parent particle, the parent particle decays, among other possible
states, into the final states containing neutrinos 
\eqarr{
&&\sum_{j}\int\![d^3\bP_F]\,
\chi_{\alpha j}(\ms{\{\bP_F\}})\,e^{-iE_{Fij}t}
\ket{X,\nu_j;\ms{\{\bP_F\}}}~,
\\
&&\sum_{j}\int\![d^3\bP_F]\,
\chi_{\alpha j}(\ms{\{\bP_F\}})\,e^{-iE_{Fij}t}
\ket{X,\bar{l}_\alpha,\nu_j;\ms{\{\bP_F\}}}~,
}
where $E_{Fj}$ is total energy of the final state, $\bP_F$
denotes the final total 3-momentum and $\{\bP_F\}$ is the set of the final momenta.

A calculation based on the Wigner-Weisskopf approximation\,\cite{intrinsic,WW}
yields 
\eqarr{
\chi_{\alpha j}(\ms{\{\bPF\}})&=&
U^*_{\alpha j}F_{\alpha j}\psi_I(\bPF)
\cr
\label{F:def}
U^*_{\alpha j}F_{\alpha j}&\equiv&
N_{\alpha j}^{-1/2}\frac{U^*_{\alpha j}f\cM_{\alpha j}}
{\displaystyle\Delta E_{\alpha j}-i\Lambda}
\,,
}
where we used the shorthands $\Delta E_{\alpha j}\equiv E_I-E_{Fj}$,
$N_{\alpha j}\equiv(2\pi)^{-3} 2E_I[2E_F]$
and $\Lambda\equiv
\frac{\Gamma}{2\gamma}$, with $\Gamma$ being the total decay width and
$\gamma=\frac{E_I}{M_I}$ the Lorentz factor; $[2E_F]=\prod_{i}(2\pi)^{3}2E_{Fi}$ is
the product of the final energies;
$I=l^+_\alpha$ for type A decay.
The factor $\Delta E_{\alpha j}-i\Lambda$ in the denominator is responsible for
ensuring approximate energy conservation since
\eq{
\label{DeltaE=0}
\frac{\Lambda}
{|\Delta E-i\Lambda|^2}
\stackrel{\Lambda\to 0}{\longrightarrow}
\pi\,\delta(\Delta E)
\,.
}
The factor $\cM_{\alpha j}$ is the invariant amplitude for the momentum
defined process, with the mixing factor $U^*_{\alpha j}$ factored
out\,\cite{endnote3}, while
$f$ is a form factor necessary to regularize the expressions for large
momenta\,\cite{intrinsic}.
To compute the flavor violation probabilities, however, the form factor is not
necessary\,\cite{intrinsic}.
Recall that the production probability density for each mass definite channel is
given by $|\chi_{\alpha j}|^2$.

Taking Eq.\,\eqref{F:def} into account, the creation probability for both type A and
B decays are
\eq{
\label{FvioAB:2}
\mathcal{P}_{l_\alpha\nu_\beta}(t)=
\int\! d^3\bp\,|\psi_I(\bp)|^2\!
\int\![d^3\bP_F]\delta^3(\bP_F-\bp)\,
\Big|\sum_j
U_{\beta j}e^{-iE_{\nu_j}t}U^{\dag}_{j\alpha}F_{\alpha j}
\Big|^2
\,,
}
provided that we use the appropriate quantity $F_{\alpha j}$.
We see the exponential $e^{-iE_{\nu_j}t}$ is responsible for the neutrino
oscillation phenomenon\,\cite{intrinsic,flavorLee}.
Despite of that, we can see the sum of probabilities
\eq{
\label{P:tot:G}
\sum_{\beta=e,\mu,\tau}\mathcal{P}_{l_\alpha\nu_\beta}(t)=
\int\! d^3\bp\,|\psi_I(\bp)|^2\!
\int\![d^3\bP_F]\delta^3(\bP_F-\bp)\,
\sum_{j}|U_{\alpha j}|^2|F_{\alpha j}|^2
\,
}
is time independent and will be shown to be approximately equal to the total
probability of neutrino production.
Ultimately, however, we will be interested in the flavor violation at creation,
\textit{i.e.},
at times $t$ that satisfy $1/\Gamma \ll t\ll L_{\rm osc}$. We will assume henceforth
that the time in question satisfies such regime and references to time will be
suppressed.

To check the correct normalization explicitly, we consider
the probability in Eq.\,\eqref{FvioAB:2} for the type B decay $I\to
\bar{l}_\alpha\nu_\beta X$ and take the limits: (a) massless neutrinos, $m_j\to
0$, (b) small width \eqref{DeltaE=0} and (c) parent particle at rest and
with sufficiently small momentum uncertainty, \textit{i.e.}, 
$|\psi_I(\bp)|^2$ is only appreciable around $\bp\approx 0$, within a size
$|\bp|\lesssim \sigma_p$, for which the rest of the integrand varies very slowly. We
obtain
\eq{
\label{FvioAB:m=0}
\mathcal{P}\ms{(I\to \bar{l}_\alpha\nu_\beta^\zero X)}
=
\frac{\delta_{\alpha\beta}}{\Gamma}
\frac{(2\pi)^4}{2M_I}
\int\!\frac{[d^3\bPF]}{[2E_F]}\delta^4(P_F-P_I)\,
|\cM_{\alpha}|^2
=\delta_{\alpha\beta}
\frac{\Gamma\ms{(I\to \bar{l}_\alpha\nu_\alpha^\zero X)}}{\Gamma}
\,,
}
where $P_I=(M_I,\bs{0})$ and Eq.\,\eqref{FvioAB:m=0} coincides with the branching
ratio for $\alpha=\beta$.
The notation $\nu^\zero$ indicates we are considering the respective neutrino, or
combination of neutrinos, massless.
Notice Eq.\,\eqref{FvioAB:m=0} is Lorentz invariant and considering the parent
particle at rest in restriction (c) above is not essential.
Notice, the order of the limits is important and we are taking (a) before (b); if
we take (b) before (a) we obtain the incoherent limit discussed in
Sec.\,\ref{sec:discussion}.
It is also important to emphasize that Eq.\,\eqref{FvioAB:m=0} is flavor diagonal 
($\sim\delta_{\alpha\beta}$), thus confirming that neutrino flavor is
a well defined concept for massless neutrinos in the SM\,\cite{intrinsic,flavorLee}.
The same analysis can be performed for type A decays.

For comparison we can calculate, under the same conditions but finite neutrino mass,
the probability for the mass definite channel
\eq{
\label{FvioAB:mj}
\mathcal{P}\ms{(I\to \bar{l}_\alpha\nu_j X)}
=
\frac{|U_{\alpha j}|^2}{\Gamma}\frac{(2\pi)^4}{2M_I}
\int\!\frac{[d^3\bPF]}{[2E_F]}\delta^4(P_F-P_I)\,
|\cM_{\alpha j}|^2
=|U_{\alpha j}|^2\frac{\Gamma\ms{(I\to \bar{l}_\alpha\nu_j X)}}{\Gamma}
\,.
}
We can define the total probability
\eq{
\label{P:tot}
\mathcal{P}_{\rm tot}\ms{(I\to \bar{l}_\alpha\nu_\alpha X)}
\equiv
\sum_j\mathcal{P}\ms{(I\to \bar{l}_\alpha\nu_j X)}
\,.
}
In special, when there is only one channel involving one charged lepton such as
$\mu^+\to e^+\nu_e\bnu_\mu$, the probability above is unity.
The probability \eqref{P:tot} is the usual production probability when neutrinos are
not detected\,\cite{shrock}. 
When the emission spectrum is considered for the charged lepton, the
distortion of the endpoint compared to the massless neutrino case gives us the
effective absolute neutrino masses\,\cite{absnumass,shrock}.
In general, however, we can approximate
\eq{
\mathcal{P}_{\rm tot}\ms{(I\to \bar{l}_\alpha\nu_\alpha X)}
\approx
\mathcal{P}\ms{(I\to\bar{l}_\alpha\nu_\alpha^\zero X)}\,.
}
The difference is negligible [$\sim (\Delta m/E_\nu)^2$] and goes to zero for
massless neutrinos.
Then, the sum of probabilities in Eq.\,\eqref{P:tot:G} is equal to Eq.\,\eqref{P:tot}
within the small width approximation.
If we divide Eq.\,\eqref{P:tot:G} by Eq.\,\eqref{P:tot} we obtain
\eq{
\label{P:tot:1}
\sum_{\beta=e,\mu,\tau}\tilde{\mathcal{P}}_{l_\alpha\nu_\beta}
=1\,,
}
where
$\tilde{\mathcal{P}}_{l_\alpha\nu_\beta}=\mathcal{P}_{l_\alpha\nu_\beta}/(\mathcal{P
}_{l_\alpha\nu_\alpha})_{\rm tot}$.

To proceed further in the analysis of Eq.\,\eqref{FvioAB:2}, we can consider some
approximations.
Due to Eq.\,\eqref{DeltaE=0}, approximate energy conservation holds and the
amount of violation is of the order of $\Gamma$.
Then, around the energy conserving values, due to simple kinematics,
neutrino masses are negligible in the terms $\cM_{\alpha j}$ and $N_{\alpha j}$
which allows us to approximate them to
\eq{
\label{mnu=0}
\cM_{\alpha}\equiv (\cM_{\alpha j})_{m_j\rightarrow 0}^{\rm EC}~,~~
N_{\alpha}\equiv (N_{\alpha j})_{m_j\rightarrow 0}
\,,
}
where EC denotes that energy conservation is strictly assumed\,\cite{endnote1}.
In other words, neutrinos are always produced ultra-relativistic and neutrino mass
is not the leading term within any multiplicative factor.
Approximations in Eq.\,\eqref{mnu=0} allows us to write Eq.\,\eqref{FvioAB:2} as
\eq{
\label{FvioAB:3}
\mathcal{P}_{l_\alpha\nu_\beta}=
\int\!\frac{d^3\bp}{2E_I}\,|\psi_I(\bp)|^2\!
\int\!\frac{[d^3\bP_F]}{[2E_F]}(2\pi)^3\delta^3(\bP_F-\bp)\,
|\cM_\alpha|^2
\Big|\sum_j
\frac{U_{\beta j}U^{\dag}_{j\alpha}}
{\Delta E_{\alpha j}-i\Lambda}
\Big|^2
\,.
}
Moreover, within the approximations of Eq.\,\eqref{mnu=0}, it can be shown (appendix
\ref{ap:psi=1}) that the integral in $\bp$ can be decoupled and we can rewrite 
Eq.\,\eqref{FvioAB:3} as 
\eq{
\label{FvioAB:4}
\mathcal{P}_{l_\alpha\nu_\beta}=
\frac{1}{2M_I}
\int\!\frac{[d^3\bP_F]}{[2E_F]}(2\pi)^3\delta^3(\bP_F)\,
|\cM_\alpha|^2
\Big|\sum_j
\frac{U_{\beta j}U^{\dag}_{j\alpha}}
{\Delta E_{\alpha j}-i\Lambda}
\Big|^2
\,,
}
where all the quantities are calculated in the rest frame of the parent
particle $I$.

We can also make the flavor violating contributions explicit by rewriting the term
inside the square modulus in Eq.\,\eqref{FvioAB:2} as
\eq{
\sum_{j=1}^{3}U_{\alpha j}U^*_{\beta j}F_{\alpha j}=
\delta_{\alpha\beta}F_{\alpha 1}+\sum_{j=2}^{3}U_{\alpha j}U^*_{\beta j}\Delta
F_{\alpha j}
\,,
}
where $\Delta F_{\alpha j}\equiv F_{\alpha j}-F_{\alpha 1}$.
Thus the square modulus becomes
\eq{
\label{sum23}
|\sum_{j=1}^{3}U_{\alpha j}U^*_{\beta j}F_{\alpha j}|^2=
\delta_{\alpha\beta}|F_{\alpha 1}|^2+
\delta_{\alpha\beta}2\mathrm{Re}
\Big[F_{\alpha 1}^*\sum_{j=2}^{3}U_{\alpha j}U^*_{\beta j}\Delta F_{\alpha j}
\Big]
+
\Big|\sum_{j=2}^{3}U_{\alpha j}U^*_{\beta j}\Delta F_{\alpha j}\Big|^2
\,.
}
We recognize that only the last term of Eq.\,\eqref{sum23} is flavor
non-diagonal.
If approximations \eqref{mnu=0} are considered, we can write
$\Delta F_{\alpha j}$ as
\eq{
\label{DeltaF:approx}
\Delta F_{\alpha j}\approx
\frac{\cM_\alpha}{N_\alpha}
\frac{(\Delta E_{\nu_j})}
{(\Delta E_{\alpha j}-i\Lambda)(\Delta E_{\alpha 1}-i\Lambda)}
\,,\quad j=2 \text{ or } 3,
}
where $\Delta E_{\nu_j}\equiv E_{\nu_j}-E_{\nu_1}$.

Specializing to $\alpha\neq\beta$, under the approximation of one
dominant contribution (two families), the initial creation probability yields
\eq{
\label{Pvio:2f}
\mathcal{P}_{l_\alpha\nu_\beta}=
|U_{\alpha j}U_{\beta j}^*|^2
\int\! d^3\!\bp\,|\psi_I(\bp)|^2\!
\int\![d^3\bPF]\,
|\Delta F_{\alpha j}|^2
\,,\quad j=2 \text{ or } 3.
}

\section{Neutrino flavor violation in muon decay}
\label{sec:mudecay}

We will calculate here the intrinsic neutrino flavor violation probability for
muon decay $\mu^+\rightarrow \bnu_\alpha e^+\nu_\beta$. Both types A and B of
violation will be possible, \textit{i.e.}, (A) $\alpha \neq \mu$ and (B) $\beta \neq
e$.
The calculations developed here can be generally adapted to consider the flavor
violation probability for any three body decay with neutrinos emerging as final
states.

In terms of mass eigenstates, the decay $\mu^+\rightarrow \bnu_\alpha
e^+\nu_\beta$ should be viewed as a coherent superposition of the 6 channels
$\mu^+\rightarrow \bnu_i e^+\nu_j$, $i,j=1,2,3$.
The calculation for each mass eigenstate channel with fixed momenta follows the
momentum convention $\mu^+(p)\rightarrow \bnu_i(k)e^+(q)\nu_j(k')$.

Following the description of Sec.\,\ref{sec:decay}, the initial muon state
\eq{
\ket{\mu}=\int d^3\bp\, \psi(\bp)\ket{\mu(\bp)}\,,
}
after a time $t\gg 1/\Gamma$, decays into the state
\eq{
\label{mu:enunu}
\sum_{ij}\int\! d^3\bq d^3\bk d^3\bk'\,
\chi_{\mu i,ej}(\bq,\bk,\bk')\,e^{-iE_{Fij}t}
\ket{e(\bq)\bnu_i(\bk)\nu_j(\bk')}~,
}
where $E_{Fij}=E_e(\bq)+E_{\bnu_i}(\bk)+E_{\nu_j}(\bk')$.
Considering Eq.\,\eqref{F:def}, $\chi_{\mu i,ej}(\bq,\bk,\bk')=
U_{\mu i}U^*_{ej} F_{\mu i,ej}\psi\ms{(\bPF)}$ and the initial
creation probability for $\mu^+\rightarrow \bnu_\alpha e^+\nu_\beta$ is then
\eq{
\label{P:mu:1}
\mathcal{P}_{\to\bnu_\alpha\nu_\beta}=
\int\! d^3\bp\,|\psi(\bp)|^2\!
\int\!d^3\bq d^3\bk d^3\bk'\delta^3(\bP_F-\bp)\,
\Big|\sum_{ij}
U_{\mu i}U^*_{\alpha i}
U_{\beta j}U^*_{ej}
F_{\mu i,ej}
\Big|^2
\,,
}
where $\bP_F=\bq+\bk+\bk'$ and
\eqarr{
U_{\mu i}U^*_{ej}F_{\mu i,ej}&=&
N_{ij}^{-1/2}\frac{f\cM_{ij}}
{\displaystyle\Delta E_{ij}-i\Lambda}
\,,\\
\cM_{ij}&=&
\tilde{G}\,
U_{\mu i}U^*_{ej}\tilde{\cM}_{ij}\,,
}
where $\tilde{G}\equiv 2\sqrt{2}G_F$, $\Delta
E_{ij}=E_\mu(\bp)-E_e(\bq)-E_{\bnu_i}(\bk)-E_{\nu_j}(\bk')$ and
$\tilde{\cM}_{ij}$ is given in Eq.\,\eqref{tM:mu}.
The probability amplitude squared for unpolarized muon is
\eq{
\mn{\frac{1}{2}}\sum_{\rm spins}
|\tilde{\cM}_{ij}|^2=
8(p\ponto k')(q\ponto k)\,.
}
By using the approximation \eqref{mnu=0}, Eq.\,\eqref{P:mu:1} becomes
\eq{
\label{P:mu:2}
\mathcal{P}_{\to\bnu_\alpha\nu_\beta}=
\frac{8\tilde{G}^2}{2M_\mu}\!
\int\!\frac{[d^3\bPF]}{[2E_F]}
(2\pi)^3\delta^3(\bP_F)\,
(p\ponto k')(q\ponto k)
\Big|\sum_{ij}
\frac{
U_{\mu i}U^*_{\alpha i}
U_{\beta j}U^*_{ej}
}{
\displaystyle\Delta E_{ij}-i\Lambda
}
\Big|^2
\,,
}
where the muon is at rest.

After following the calculations described in appendix \ref{ap:calc}, we can
obtain the flavor violation probabilities for type A and B.

The type A neutrino flavor violation probability, within two families
approximation \eqref{Pvio:2f}, is given by
\eq{
\label{P:mu:A:f}
\mathcal{P}\ms{(\mu^+\!\to\!\bnu_\alpha e^+\nu_e^{\mt{(0)}})}=
2|U_{\mu i}U^*_{\alpha i}|^2
\int_0^{1}dx_\nu
\cS_{\bnu_\mu}\ms{(x_\nu)}
\rho_{i1}\ms{(x_\nu)}\,,
\quad i=2\text{ or }3,
}
where $x_\nu\equiv E_{\bnu}/W_\bnu$, $W_\bnu\sim M_\mu/2$ is the maximum antineutrino
energy ($0\le x_\nu\le 1$) and $\cS_{\bnu_\mu}$ is the emission probability (energy
spectrum) for $\bnu_\mu$ in the ordinary channel $\mu^+\to\bnu_\mu
e^+\nu_e$; see Eq.\,\eqref{dist:numu}.
The flavor violating spectrum is then modified by the mixing factor in front of the
integral in Eq.\,\eqref{P:mu:A:f} and 
\eqarr{
\label{rho:1}
\rho_{ij}(x)
&=&
\frac{(\Delta E_{\bnu_i})^2}
{(\Delta E_{\bnu_i})^2+4\Lambda^2}\, ,
\\
\label{rho:2}
&=&
\frac{\omega^2_{ij}}{x^2+\omega^2_{ij}+\bar{a}^2_{ij}+b_{ij}(x)}\,,
}
which depends on $a_i\equiv m_i/W_\bnu$, $\bar{a}_{ij}=\frac{1}{2}(a_i+a_j)$,
\eqarr{
\label{delta:def}
\omega_{ij}&=&\frac{m^2_i-m^2_j}{2\Gamma\,W_\nu}\,,
\\
b_{ij}(x)&=&\frac{1}{2}\Big[\sqrt{x^2+a^2_i}\sqrt{x^2+a^2_j}-x^2-a_ia_j\,\Big]\,.
}
The monotonically increasing function $b_{ij}(x)$ has range $[0,(\Delta
a)^2/4]$, where $\Delta a\equiv a_i-a_j$; thus $\bar{a}^2_{ij}+b_{ij}\le
\frac{1}{2}(a^2_i+a^2_j)\approx 4m^2_\nu/M^2_\mu$ which is negligible. Such function
is similar to the function $f$ found in Refs.\,\cite{ccn:no12,bernardini} in another
context.

The distribution function $\cS_{\bnu_\mu}\ms{(x_\nu)}$, at tree
level and null electron mass, can be found in appendix \ref{ap:mudecay}.
It is a monotonically increasing function going from 0 to a maximum.
On the other hand, $\rho_{ij}$ in Eq.\,\eqref{rho:1} increases for small $x$.
Within these approximations, the expression \eqref{P:mu:A:f} can be integrated and
we obtain, at leading order\,\eqref{int:x23},
\eq{
\label{P:mu:A:int}
\mathcal{P}\ms{(\mu^+\!\to\!\bnu_\alpha e^+\nu_e^{\mt{(0)}})}=
8|U_{\mu i}U^*_{\alpha i}|^2
\omega^2_{i1}\,,\quad i=2\text{ or }3.
}
We have typically
\eqarr{
\label{P:mu:A:int2}
\mathcal{P}\ms{(\mu^+\!\to\!\bnu_e e^+\nu_e^{\mt{(0)}})}
&=&
8|U_{\mu 2}U^*_{e 2}|^2\omega^2_{21}
\sim 0.8\,\omega^2_{12}\sim 5\times 10^{-6}\,,
\\
\label{P:mu:A:int3}
\mathcal{P}\ms{(\mu^+\!\to\!\bnu_\tau e^+\nu_e^{\mt{(0)}})}
&=&
8|U_{\mu 3}U^*_{\tau 3}|^2\omega^2_{31}
\sim 2\,\omega^2_{23}\sim 10^{-2}
\,,
}
where we have used $|U_{\mu 2}U^*_{e 2}|^2\sim |U_{\mu 2}U^*_{\tau 2}|^2\sim
0.1$, $|U_{\mu 3}U^*_{\tau 3}|^2\sim 0.24$\,\cite{vissani:review} and we have assumed
$|U_{\mu 3}U^*_{e3}|^2/|U_{\mu
2}U^*_{e2}|^2\ll (\Delta m^2_{12})^2/(\Delta m^2_{23})^2\sim 10^{-3}$ in
Eq.\,\eqref{P:mu:A:int2}.
If $|U_{\mu 3}U^*_{e3}|^2\sim 10^{-3}$, we have to consider, instead of
Eq.\,\eqref{P:mu:A:int2},
\eq{
\label{P:mu:A:int2'}
\mathcal{P}\ms{(\mu^+\!\to\!\bnu_e e^+\nu_e^{\mt{(0)}})}=
8|U_{\mu 3}U^*_{e 3}|^2 \omega^2_{31}
\sim 0.8\times 10^{-2}\,\omega^2_{23}\sim 5\times 10^{-5}\,.
}
We made use of the experimental values
\eq{
\omega^2_{12}=\Big(\frac{\Delta m^2_{12}}{2W_\nu\Gamma_\mu}\Big)^2
\approx 5.8\times 10^{-6}\,,
\quad
\omega^2_{23}=\Big(\frac{\Delta m^2_{23}}{2W_\nu\Gamma_\mu}\Big)^2
\approx 5.7\times 10^{-3}\,,
}
with
$|\Delta m^2_{12}|\approx 7.6\times 10^{-5}\rm
eV^2$ and $|\Delta m^2_{23}|\approx 2.4\times 10^{-3}\rm
eV^2$\,\cite{vissani:review}.
Additionally, if we had used Eq.\,\eqref{int:x23} for $\omega_{23}$, we would have
obtained a value 18\% larger in Eq.\,\eqref{P:mu:A:int3}.

Analogously, the type B neutrino flavor violation probability is given by
\eq{
\label{P:mu:B:f}
\mathcal{P}\ms{(\mu^+\!\to\!\bnu_\mu^{\mt{(0)}}e^+\nu_\beta)}=
2|U_{\beta j}U^*_{ej}|^2
\int_0^{1}dx'_\nu
\cS_{\nue}\ms{(x'_\nu)}
\rho_{j1}\ms{(x'_\nu)}\,,
}
where $x_\nu\equiv E_{\nu}/W_\nu$, $W_\nu\sim M_\mu/2$, and the remaining functions
are the same as in Eq.\,\eqref{P:mu:A:f}; 
the energy distribution $\cS_{\nue}\ms{(x'_\nu)}$ is different and can be found in
Eq.\,\eqref{dist:nue}.
After integration we obtain, at leading order,
\eq{
\label{P:mu:B:int}
\mathcal{P}\ms{(\mu^+\!\to\!\bnu_\mu^{\mt{(0)}}e^+\nu_\beta)}=
12|U_{\beta j}U^*_{ej}|^2
\omega^2_{j1}\,,\quad j=2\text{ or }3.
}
The typical values for $\beta=\mu$ and $\beta=\tau$ are
\eqarr{
\label{P:mu:B:int2}
\mathcal{P}\ms{(\mu^+\!\to\!\bnu_\mu^{\mt{(0)}}e^+\nu_\mu)}
&=&
12|U_{\mu 2}U^*_{e2}|^2\omega^2_{21}
\sim 1.2\, \omega^2_{12}\sim 7\times 10^{-6}
\,,
\\
\label{P:mu:B:int3}
\mathcal{P}\ms{(\mu^+\!\to\!\bnu_\mu^{\mt{(0)}}e^+\nu_\tau)}
&=&
12|U_{\tau 2}U^*_{e2}|^2\omega^2_{21}
\sim 1.2\,\omega^2_{12}\sim 7\times 10^{-6}
\,,
}
where we used $|U_{\tau 2}U^*_{e2}|^2\sim 0.1$, in addition to the values already
used after Eq.\,\eqref{P:mu:A:int3}. 
If $|U_{\mu 3}U^*_{e3}|^2\sim 10^{-3}$, we have to consider, instead of
Eq.\,\eqref{P:mu:B:int3},
\eq{
\label{P:mu:B:int3'}
\mathcal{P}\ms{(\mu^+\!\to\!\bnu_\mu^{\mt{(0)}}e^+\nu_\tau)}
=
12|U_{\tau 3}U^*_{e3}|^2\omega^2_{31}
\sim 1.2\times 10^{-2}\omega^2_{23}\sim 7\times 10^{-5}\,.
}
In addition, if we had used Eq.\,\eqref{int:x23} for $\omega_{23}$, we would
have obtained a value 25\% larger in Eq.\,\eqref{P:mu:B:int3}.

The difference in the factor $3/2$ between type A violation probability in
Eq.\,\eqref{P:mu:A:int} and type B violation probability in
Eq.\,\eqref{P:mu:B:int} can be understood by analyzing the overlap between
the function $\rho_{ij}$ in Eq.\,\eqref{rho:1} and the emission  spectra of the
would-be $\bnu_\mu$ (type A), $\cS_{\bnu_\mu}\ms{(x_\nu)}$, and the would-be
$\nu_e$ (type B), $\cS_{\nue}\ms{(x'_\nu)}$.
These distributions can be found in appendix \ref{ap:mudecay}.
Considering that $\rho_{ij}(x)$ is larger for small $x$,
the overlap with $\cS_{\nue}\ms{(x'_\nu)}$ is larger than with
$\cS_{\bnu_\mu}\ms{(x_\nu)}$ because the former is centered around a smaller value.

It is also important to emphasize that from Eqs.\,\eqref{P:mu:A:f} and
\eqref{P:mu:B:f} we can extract the flavor violation emission spectrum for neutrinos
in type A violation, $\mu^+\to \bnu_\alpha e^+\nu_e$, $\alpha\neq \mu$, or type
B violation, $\mu^+\to \bnu_\mu e^+\nu_\beta$, $\beta\neq e$. 
The emission spectrum of the wrong-type neutrinos relative to the spectrum of the
correct-type neutrinos is given by the function $\rho(x)$ in Eq.\,\eqref{rho:2}.
Such function increases for smaller neutrino energies, which implies that the
relative effect of neutrino flavor violation is larger for neutrinos produced in the
low energy part of the spectrum.

\section{Equal-energy neutrino flavor states}
\label{sec:DeltaE=0}

By assuming the usual neutrino flavor states \eqref{nubeta}, we have found in
Sec.\,\ref{sec:mudecay} a flavor violation as large as 1\% in the channel $\mu^+\to
\bnu_\tau e^+\nu_e$ relative to the dominant channel $\mu^+\to \bnu_\mu e^+\nu_e$. It
is then mandatory to check if a different definition of the neutrino flavor states
can minimize the inherent flavor indefiniteness that arises.

Qualitatively, we know from the relevant quantity $\omega_{ij}$ \eqref{delta:def}
that flavor violation was present because the size of the neutrino
energy differences $\Delta E_\nu\sim \Delta m^2_i/2E_\nu$ is comparable to the
intrinsic energy uncertainty of the creation process given by $\delta E\sim
\Gamma$. Thus, it is important to remark that although the calculations took into
account both the energy uncertainty quantified by $\Gamma$ and the momentum
uncertainty $\sigma_p$ encoded in the parent particle momentum wave function
$\psi(\bp)$, the latter did not play any relevant role in the flavor violation
probability (it could indeed taken to be zero).
The reason lies on the fact that by calculating the probability to detect the 
neutrino flavor states \eqref{nubeta}, we are implicitly summing over neutrino mass
eigenstates with the same momentum.

We can define, instead, the equal-energy neutrino flavor states
\eqarr{
\label{nu:DeltaE=0}
\ket{\nu_\alpha{:}E,\hat{\bk}}
&\equiv &
U^*_{\alpha i}\ket{\nu_i(\rk_i\hat{\bk})}\,,\quad \rk_i=\mn{\sqrt{E^2-m^2_i}}\,,
\nonumber\\
\ket{\bnu_\alpha{:}E,\hat{\bk}}
&\equiv &
U_{\alpha i}\ket{\bnu_i(\rk_i\hat{\bk})}\,,\quad \rk_i=\mn{\sqrt{E^2-m^2_i}}\,.
}
The superpositions involve neutrino states with the same energy since
$E_{\nu_i}(\rk_i\hat{\bk})=E$, $i=1,2,3$. It is also necessary to adopt the
convention $E\ge m_3=\max(m_i)$.
Notice, the states \eqref{nu:DeltaE=0} do not obey the orthogonality condition
\eqref{nu:orto}.

We can recalculate the flavor violation probability of Sec.\,\eqref{sec:mudecay}
for muon decay by adopting Eq.\,\eqref{nu:DeltaE=0}. For the type A violation
$\mu^+\to \bnu_\alpha e\nu_e^\zero$, by projecting the state \eqref{mu:enunu} into
$\bra{\bnu_\alpha{:}E,\hat{\bk}}\bra{\nu_e^\zero(\bk')}\bra{e(\bq)}$, 
Eq.\,\eqref{P:mu:1} is modified to
\eq{
\label{P:mu:DeltaE:1}
\mathcal{P}\ms{(\mu^+\!\to\!\bnu_\alpha e^+\nu_e^\zero)}
=
\int\!d^3\bq (d^3\bk)_E d^3\bk'\,
\Big|\sum_{i}
U_{\mu i}U^*_{\alpha i}
F_{\mu i}
\psi(\bq+\rk_i\hat{\bk}+\bk')
\Big|^2
\,,
}
where $\int(d^3\bk)_E$ denotes $\int_{m_3}^{\infty}dE E^2\int d\Omega_k$. (Other
choices are possible but they are not relevant as long as
approximation \eqref{mnu=0} is valid.)
We have used
\eq{
\braket{\bnu_\alpha{:}E,\hat{\bk}}
{\bar{\nu}_i(\bk')}=
U^*_{\alpha i}\,\delta^3(\bk'-\rk_i\hat{\bk})\,,\quad \rk_i=\mn{\sqrt{E^2-m^2_i}}\,.
}

We then use approximation \eqref{mnu=0} to rewrite
\eq{
\label{P:mu:DeltaE:2}
\mathcal{P}\ms{(\mu^+\!\to\!\bnu_\alpha e^+\nu_e^\zero)}
=
\int\!\!d^3\bp d^3\bq (d^3\bk)_1 d^3\bk'\,
\delta^3(\bp-\bP_{F_1})
\frac{|\cM_{\mu}|^2}{N^2_{\mu}}
\Big|\sum_{i}
\frac{U_{\mu i}U^*_{\alpha i}\,\psi_i}
{\Delta E_{\mu i}-i\Lambda}
\Big|^2
\,,~~
}
where $\bP_{F_1}=\bq+\rk_1\hat{\bk}+\bk'$,
$\Delta \rk_i=\rk_i-\rk_1$, $(d^3\bk)_1=d^3(\rk_1\hat{\bk})$ and
\eqarr{
\label{psi:i}
\psi_i&\equiv&\psi\ms{(\bp+\Delta\rk_i\hat{\bk})}\,,
\\
\label{DeltaE:i}
\Delta E_{\mu i}&\equiv&
E_\mu\ms{(\bp+\Delta\rk_i\hat{\bk})}-E_e(\bq)-E-E_\nu(\bk')\,.
}
In the two families approximation, Eqs.\,\eqref{psi:i} and
\eqref{DeltaE:i} contribute as
\eqarr{
\label{Delta:psi,E}
\frac{\psi_i}{\Delta E_{\mu i}-i\Lambda}-\frac{\psi_1}{\Delta E_{\mu i}-i\Lambda}
&\approx&
\frac{\Delta\psi_i}{\Delta E_{\mu 1}-i\Lambda}+
\psi_1\Delta\Big(\frac{1}{\Delta E_{\mu i}-i\Lambda}\Big)\,,
\\
\label{Delta:psi}
\Delta\psi_i=\psi_i-\psi_1 &\approx&
\Delta\rk_i\,\hat{\bk}\ponto\frac{\partial }{\partial\bp}\psi(\bp)\,,
\\
\label{Delta:E}
\Delta\Big(\frac{1}{\Delta E_{\mu i}-i\Lambda}\Big)
&=&
\frac{E_\mu\ms{(\bp)}-E_\mu\ms{(\bp+\Delta\rk_i\hat{\bk})}}
{(\Delta E_{\mu 1}-i\Lambda)(\Delta E_{\mu i}-i\Lambda)}
\,.
}

Let us now calculate the flavor violation probability \eqref{P:mu:DeltaE:2} in two
regimes: 
(i) $\sigma_x\gg 1/\sqrt{\Gamma_\mu M_\mu}$
~\big[$|\Delta\psi_i|$ dominates over
$|E_\mu\ms{(\bp)}-E_\mu\ms{(\bp+\Delta\rk_i\hat{\bk})}|$\big]
and 
(ii) $\sigma_x\ll 1/\sqrt{\Gamma_\mu M_\mu}$ ~
\big[$|E_\mu\ms{(\bp)}-E_\mu\ms{(\bp+\Delta\rk_i\hat{\bk})}|$
dominates over $|\Delta\psi_i|$\big].
The details are shown in appendix \ref{ap:calc}.

\begin{itemize}
\item[(i)]
{$\sigma_x\gg 1/\sqrt{\Gamma_\mu M_\mu}$}
\eqarr{
\label{P:mu:DeltaE:i1}
\mathcal{P}\ms{(\mu^+\!\to\!\bnu_\alpha e^+\nu_e^\zero)}
&=&
\ml{\frac{1}{2}}|U_{\mu i}U^*_{\alpha i}|^2
(\sigma_x W_\nu)^2
\!\int_0^1\!\! dx\,S_{\bnu_\mu}(x)
\Big(\!\sqrt{x^2+|\Delta a^2_i|}-x\Big)^2
\\
\label{P:mu:DeltaE:i2}
&=&
\ml{\frac{1}{2}}|U_{\mu i}U^*_{\alpha i}|^2
\Big(\frac{\Delta m^2_i\sigma_x}{W_\nu}\Big)^2
\,,\quad i=2,3;
}

\item[(ii)]
{$\sigma_x\ll 1/\sqrt{\Gamma_\mu M_\mu}$}
\eqarr{
\label{P:mu:DeltaE:ii1}
\mathcal{P}\ms{(\mu^+\!\to\!\bnu_\alpha e^+\nu_e^\zero)}
&=&
|U_{\mu i}U^*_{\alpha i}|^2
\Big(\frac{\sigma_pW_\nu}{M_\mu\Gamma}\Big)^2
\!\int_0^1\!\! dx\,S_{\bnu_\mu}(x)
\Big(\!\sqrt{x^2+|\Delta a^2_i|}-x\Big)^2
\\
\label{P:mu:DeltaE:ii2}
&=&
4|U_{\mu i}U^*_{\alpha i}|^2
\omega^2_{i1}\frac{\sigma^2_p}{M^2_\mu}
\,,\quad i=2,3.
}
\end{itemize}
We have used $\sigma^2_x=\int d^3\bp|\partial\psi(\bp)/\partial \bp|^2$ and
$\sigma^2_p=\int d^3\bp|\bp|^2|\psi(\bp)|^2$ for the position and momentum
uncertainties of the parent particle, respectively. (Assuming the mean position and
momentum are null.) The shorthand $\Delta a^2_i=a^2_i-a^2_1$ was also used.

We can compare the contributions of Eqs.\,\eqref{P:mu:DeltaE:i2} and
\eqref{P:mu:DeltaE:ii2} by using $\sigma_p\sim 1/2\sigma_x$ and
inserting typical values for $\alpha=\tau$ and $i=3$:
\eqarr{
\label{P:mu:DeltaE:i3}
\text{(i)}\quad \mathcal{P}\ms{(\mu^+\!\to\!\bnu_\tau e^+\nu_e^\zero)}
&\sim&
6\times 10^{-13}\Big(\frac{\sigma_x}{1\rm cm}\Big)^2\,,
\\
\label{P:mu:DeltaE:ii3}
\text{(ii)}\quad \mathcal{P}\ms{(\mu^+\!\to\!\bnu_\tau e^+\nu_e^\zero)}
&\sim&
6\times 10^{-29}\Big(\frac{1\rm cm}{\sigma_x}\Big)^2\,.
}
We conclude that both contributions are negligible unless $\sigma_x$ is extremely
large or small compared to 1cm. In fact, Eqs.\,\eqref{P:mu:DeltaE:i2} and
\eqref{P:mu:DeltaE:ii2} have opposite behaviors as functions of $\sigma_x$; they are
comparable when $\Gamma\sigma_x\sim \sigma_p/M_\mu$ or
\eq{
\sigma_x\sim \frac{1}{\sqrt{\Gamma M_\mu}}\sim 1\mu m\,,\quad 
(\sigma_x\sim 1/\sigma_p).
}
In that case both contributions of Eqs.\,\eqref{P:mu:DeltaE:i3} and
\eqref{P:mu:DeltaE:ii3} are of the order of $10^{-21}$.

It is important to remark that with the typical values of
Eqs.\,\eqref{P:mu:DeltaE:i3} and \eqref{P:mu:DeltaE:ii3}, approximation
\eqref{mnu=0} can not, in general, be applied to Eq.\,\eqref{P:mu:DeltaE:1} to
obtain Eq.\,\eqref{P:mu:DeltaE:2}, because we might be neglecting contributions
of the order of
\eq{
\frac{\Delta m^2_{32}}{2E^2_\nu}\sim
10^{-17}\Big(\frac{10\rm MeV}{E_\nu}\Big)^2\,.
}
Therefore, to properly calculate the flavor violation probabilities for the states
\eqref{nu:DeltaE=0}, it is necessary to perform the full calculation of 
Eq.\,\eqref{P:mu:DeltaE:1}.
Nevertheless, the calculations of this section show that the flavor violation
probability is negligible for the states \eqref{nu:DeltaE=0}.

\section{Discussions and Conclusions}
\label{sec:discussion}

{}

We have calculated the flavor violating creation probability for neutrinos
produced through decay, using the full QFT formalism within the Wigner-Weisskopf
approximation, assuming that the usual neutrino flavor states
\eqref{nubeta} describe appropriately the neutrinos produced in nature.
The neutrino flavor violation occurs at creation and no propagation distance is
necessary except for the lifetime of the parent particle. 
The calculations in this paper differ from previous QFT treatments
(see Ref.\,\cite{beuthe}) in the focus on the intrinsic flavor violation, with the 
considerations of the decay width and the realistic interactions responsible for
neutrino creation. We also managed to calculate the flavor violating neutrino
emission spectrum.

For the particular case of muon decay, the amount of flavor
violation might be surprisingly large as 1\% in the channel 
$\mu^+\to \bnu_\tau e^+\nu_e$ compared to the ordinary channel $\mu^+\to \bnu_\mu
e^+\nu_e$.
We have found that the amount of flavor violation in such channel is much larger
than other flavor violating channels because the $\nu_\mu$--\,$\nu_\tau$ mixing is
large and the relevant mass difference $\Delta m^2_{13}\sim \Delta m^2_{23}$ is
relatively larger.
The relevant quantity is given by $\omega_{ij}\sim\Delta m^2_{ij}/2E_\nu\Gamma$ in
Eq.\,\eqref{delta:def}.
Compared to neutrinos produced in pion decay $\pi^+\to \mu^+\nu_\mu$, where the
amount of flavor violation of the same type was at most of the order of
$10^{-6}$\,\cite{intrinsic}, flavor violation in muon decay is much larger because
the decay width is much smaller, while the remaining variables such as neutrino
energy are similar.

Considering the amount of neutrino flavor violation calculated in muon decay, we
should consider the limits imposed by the CHORUS\,\cite{chorus} and
NOMAD\,\cite{nomad} experiments that was searching for the conversion
$\nu_\mu\to\nu_\tau$ but excluded such channel at the level of $10^{-4}$. However,
most of the $\nu_\mu$ beam comes from pion and kaon decay; the flavor violation
probability for neutrino states \eqref{nubeta} originating from pions is of the order
of $10^{-6}$\,\cite{intrinsic} which is not constrained by such experiments and, for
neutrinos originating from kaons, the flavor violation should be slightly smaller due
to larger decay width. On the other hand, there is a 5.6\% of contribution of
$\bnu_\mu$ coming from muon decay in CHORUS\,\cite{chorus:setting}. If the
sensitivity to detect $\bnu_\tau$ is the same as for $\nu_\tau$, the probability
estimated in Eq.\,\eqref{P:mu:A:int3} would be excluded. Moreover, if we extrapolate
our results in Eqs.\,\eqref{P:mu:B:int2} and \eqref{P:mu:B:int3} to the production of
$\bnu_e$ through the usual beta decay or the decay of long-lived heavier nucleus, we
obtain $\omega^2_{31}\gg 1$ which corresponds to the incoherent limit
\eqref{P:incoherent}. Since no flavor violation is observed in neutrinos produced in
nuclear reactors, we have to conclude that the neutrino flavor states \eqref{nubeta}
do not describe appropriately those neutrinos. Therefore, in contexts as common as
muon decay or beta decay, the equal-momentum flavor states \eqref{nubeta} are not
appropriate.

It is also important to discuss an important limit: the
\textit{incoherent limit} when $\Gamma\to 0$ (or $\omega_{ij}\to \infty$). 
Taking such limit in
Eq.\,\eqref{P:mu:A:f} or Eq.\,\eqref{P:mu:B:f} we obtain 
\eq{
\label{P:incoherent}
\mathcal{P}_{\mu\to\bnu_\alpha}
=2|U_{\mu i}U^*_{\alpha i}|^2
\quad\text{ or }\quad
\mathcal{P}_{\to e\nu_\beta}
=2|U_{\beta j}U^*_{e j}|^2
\,.
}
We easily recognize the expressions above as the incoherent limit if we resort to the
two families approximation where we would have $2|U_{\mu i}U^*_{\alpha
i}|^2=\frac{1}{2}\sin^22\theta$, where $\theta$ is the associated two-family mixing
angle. 
The results above \eqref{P:incoherent} are expected because in the incoherent limit
the vanishing energy uncertainty would destroy the quantum coherence necessary to
create the flavor states \eqref{nubeta} and each mass eigenstate neutrino would be
produced incoherently\,\cite{kayser:81}.
The same conclusion can be reached if we analyze the incoherent limit in
\eqref{FvioAB:3}: the square modulus of the sum in the last factor would be 
equivalent to the sum of the square moduli because the mixed terms would vanish due
to the lack of overlap; hence the expression would be equal to Eq.\,\eqref{P:tot:G}.
Moreover, the flavor violation calculated here is not negligible exactly because the
energy difference among the different mass eigenstates that compose the muon
neutrino, $\Delta E_{\nu_i}\sim \Delta m^2_{ij}/2E_\nu$, is comparable to the energy
uncertainty imposed by the decay width $\delta E\sim \Gamma$.

For that reason, we define equal-energy neutrino flavor states \eqref{nu:DeltaE=0}
in Sec.\,\eqref{sec:DeltaE=0} and calculate, for type A muon decay, the probability
do detect such states summed over all energies.
In this case, the correct amount of flavor violation depends on the
position (momentum) uncertainty of the parent particle $\sigma_x$ ($\sim
1/2\sigma_p$). As expected, if $\sigma_p\sim \Gamma$ the intrinsic flavor violation
in Eq.\,\eqref{P:mu:DeltaE:i2} is, except for numerical factors, identical to the
flavor violation for equal-momentum states \eqref{nubeta} calculated in
Eq.\,\eqref{P:mu:A:int3}.
However, $\sigma_p\sim \Gamma$ is equivalent to $\sigma_x\sim
\tau$ which is macroscopic for muons, \textit{i.e.}, $c\tau=659\rm m$.
Position uncertainty should be smaller than the order of 1cm, or even
much smaller ($10^{-8}\rm cm$) if the muon decays in a medium\,\cite{giunti:book}. 
(For Mössbauer neutrinos, the uncertainty would be of the order of atomic 
size\,\cite{lindner:mossbauer}.)
Even for $\sigma_x\sim 1\rm m$, we would obtain a flavor violation probability of
the order of $10^{-8}$ and thus no appreciable amount of flavor violation is
expected. 
Although, it should be remarked that such flavor violation probabilities are much
larger than the ones for indirect flavor violation processes such as
$\mu\to e\gamma$, with branching ratio $\sim10^{-50}$.
On the other hand, we can conclude that the equal-energy states \eqref{nu:DeltaE=0}
describe more accurately the neutrinos produced in muon decay than the equal-momentum
states \eqref{nu:DeltaE=0}, the description being more accurate for smaller decay
widths or longer lifetimes of the parent particle, \textit{e.g.}, for the usual beta
decay.

Analogously, in the recent controversy concerning Mössbauer neutrinos, it was shown
from a careful theoretical analysis\,\cite{lindner:mossbauer} that neutrinos
oscillate despite the tiny energy uncertainty. The reason is that the momentum
uncertainty, which can not be as small as the energy uncertainty, should be taken
into account.
In that case, equal-energy neutrino states also describe more accurately the
neutrinos propagating from source to detector, enabling flavor oscillations.

Two quantities control intrinsic neutrino flavor violation in neutrinos created
through decay: the decay width $\sim\Gamma$ (energy uncertainty) and the momentum
uncertainty $\sim\sigma_p$ of the parent particle. The former is intrinsic to the
parent particle while the latter might differ depending on the process of creation,
\textit{e.g.}, decay in vacuum as opposed to decay in a medium. From the calculations
performed we can conclude that the decay width is relevant only for detecting
equal-momentum states \eqref{nubeta} ($\sigma_p$ can be taken to be zero)
while the momentum uncertainty is relevant when equal-energy states
\eqref{nu:DeltaE=0} are considered. (The decay width entered in
Eq.\,\eqref{P:mu:DeltaE:ii2} but the suppression factor makes it usually negligible.)
Reference \onlinecite{grimus} derives exactly the quantity inside parenthesis in
Eqs.\,\eqref{P:mu:DeltaE:i2} and \eqref{P:mu:DeltaE:ii2} 
to be smaller than unity for flavor oscillations to take place
(called ACC and SFC conditions, Eqs.\,(4.2) and (4.7) in Ref.\,\onlinecite{grimus},
respectively).
We have shown here that such conditions are indeed necessary to avoid intrinsic
flavor violation and ensure initial flavor definition for equal-energy states
\eqref{nu:DeltaE=0}.
Interestingly, the condition $|\omega_{ij}|\ll 1$ \eqref{delta:def}, obtained here
 for equal-momentum states \eqref{nubeta}, is not obtained in
Ref.\,\onlinecite{grimus}.

Considering that any other choice for the neutrino flavor states other than
\eqref{nubeta} or \eqref{nu:DeltaE=0} still induces different
contributions for the distinct mass eigenstates with differences comparable
or larger than the contributions coming from $\Gamma$ or $\sigma_p$, we can
extrapolate that there is a minimum amount of flavor violation of the order of
\eq{
\label{Pvio:min}
\mathcal{P}_{l_\alpha\nu_\beta}
\gtrsim
|U_{\alpha i}U^*_{\beta i}|^2
\Big(\frac{\Delta m^2_{i1}}{2W_\nu\sigma_{*}}\Big)^2\,,
\quad
\sigma_*=\max(\Gamma,\sigma_p)\,,
}
for types A or B decays, as long as $\sigma_*\ll W_\nu$ and the quantity inside
parenthesis in Eq.\,\eqref{Pvio:min} is much smaller than unity.
The exact amount of flavor violation would depend on the details of the neutrino
state being detected.

We should also discuss two aspects of the same phenomenon in neutrino creation: (a)
flavor indefiniteness and (b) flavor violation. 
Both effects are related through the overall conservation of
probability \eqref{P:tot:1}, \textit{i.e.}, a non-null probability to detect the
``wrong''
neutrino flavor implies that the probability to detect the ``correct'' flavor should
be deficient by the same amount when compared to the usual result.
These effects are expected, at least, at the order of $(\Delta
m/E_\nu)^2$\,\cite{giunti:torino04,review,blasone:short} pointing toward the
impossibility to define the neutrino flavor in an exact manner. Such result can be
derived simply in first or second quantized formulations of flavor
oscillations\,\cite{review,ccn:no12}.
It is possible to avoid this intrinsic flavor violation\,\cite{blasone:short} by
defining an inequivalent vacuum and different flavor states\,\cite{BV}, but 
other problems appear\,\cite{giunti:fock}.
In this work and in Ref.\,\cite{intrinsic}, however, we have shown through concrete
calculations that intrinsic flavor violation probability can be much larger, of the
order of $(\Delta m^2/(2E_\nu\sigma_*))^2$\,\eqref{Pvio:min}.

The dependence of the neutrino flavor violation effect on the decay
width and momentum uncertainty brings about another possible effect: (c) the source
dependence of neutrino flavor\,\cite{giunti:torino04,kiers}. The difference between
the probabilities calculated through the equal-momentum states \eqref{nubeta} and
equal-energy states \eqref{nu:DeltaE=0} also indicates that neutrino flavor could
depend on the detection process\,\cite{kiers,grimus}. We did not pursue such effect
here, choosing to focus on idealized measurements (summation over momenta or
energies) which insured automatic probability normalization.
In any case, the calculations presented here suggest that neutrino flavor
is not universally defined and the effects might not be negligible as usually
assumed\,\cite{giunti:torino04}.
For example, for the \textit{same observable}, \textit{i.e.}, the probability to
detect the state $\ket{\nu_\mu}$ as defined in Eq.\,\eqref{nubeta}, summed over
momenta, yields the expected result at the level $10^{-6}$ in the decay $\pi^+\to
\mu^+\nu_\mu$ while the same deviates from the usual expectation by 1\% in the decay
$\mu^-\to \nu_\mu e^-\bnu_e$.
Even with a different definition of the $\nu_\mu$ flavor state, large flavor
violation would remain if we could hypothetically produce free muons (at rest) with
large position uncertainties of the order of its lifetime $\tau$ (or $\sigma_p\sim
\Gamma$).

One aspect that was not considered here was the role played by
entanglement\,\cite{glashow:no}.
However, even in that case, some level of flavor indefiniteness should be present
since the momenta and energies of the neutrino eigenstates are determined to be
distinct by the conservation of energy-momentum.
Source dependence should also occur since a neutrino flavor state produced in one
process might differ from another process by the values determined by the
conservation of energy momentum. 

Incidentally, source dependence as described here might account for the anomalies
found in the LSND\,\cite{lsnd} and MiniBoone\,\cite{miniboone} experiments.
The flavor violation probability of less than 1\% is the amount of violation
necessary to explain the LSND anomaly. 
Although the required type of flavor violation in LSND ($\bnu_\mu\to\bnu_e$) do not
match the largest probability found in this paper ($\bnu_\mu\to\bnu_\tau$), some kind
of source dependence could account for or, at least, ameliorate such anomalies
without requiring any new physics beyond the known three neutrino
families\,\cite{xing:unitvio,kopp}.
Moreover, the flavor violation effect found here agrees qualitatively with the
anomaly found in MiniBoone (in neutrino mode) since, from Eqs.\,\eqref{P:mu:A:f},
\eqref{P:mu:B:f}, \eqref{P:mu:DeltaE:i1} and \eqref{P:mu:DeltaE:ii1}, the effect is
(relatively) larger for neutrinos produced at the low energy portion of the muon
decay spectrum.

In addition, as a buy-product, if the amount of neutrino flavor violation at creation
estimated in Eqs.\,\eqref{P:mu:A:int2},\eqref{P:mu:A:int3},\eqref{P:mu:B:int2} and
\eqref{P:mu:B:int3} were detectable, we would gain an observable which is extremely
sensitive to $\theta_{13}$, \textit{i.e.}, $\mathcal{P}\ms{(\mu^+\to\bnu_e
e^+\nu_e)}$. Depending on how large is the value of $|U_{e3}|$ the contribution of
$\Delta m^2_{23}$ might dominate over $\Delta m^2_{12}$. Moreover, if $|U_{\mu
3}U^*_{e3}|\sim 10^{-4}$ the contributions from the two mass differences might be
comparable and even CP violation might be observable. Of course, to probe such
quantities it is necessary to detect $\nu_e$ in the states \eqref{nu:DeltaE=0} with
precision better than $10^{-6}$ compared to the main channel.

In conclusion, there could be neutrino flavor violation at creation for 
the muon neutrino produced through muon decay with detectable probability if
equal-momentum flavor states \eqref{nubeta} are detectable in some way.
In that case, the muon neutrino produced from pion decay might be slightly
distinct of the muon neutrino produced in muon decay; the distinction being
possibly observable. In general, however, equal-energy states \eqref{nu:DeltaE=0}
describe more appropriately the neutrinos produced through muon decay and the decay
of other long-lived nucleus, and no detectable neutrino flavor violation is expected,
unless the parent particle has uncommonly large position uncertainties.

\acknowledgments

This work is partially supported by the Brazilian agencies
FAPESP and CNPq through grants 09/11309-7 and 309455/2009-0.
The author thanks O.\,L.\,G.\,Peres for helpful discussions.


\appendix
\section{Decoupling of Eq.$\mbox{\,\eqref{FvioAB:3}}$}
\label{ap:psi=1}

We will show here that Eq.\,\eqref{FvioAB:3} within the approximation \eqref{mnu=0}
can be decoupled into a product of two factors, one of them being the decoupled
integral $\int d^3\bp|\psi_I(\bp)|$ which is unity.
In other words, we will show here that the factor
\eq{
\label{gconst:def}
g(\bp)=
\frac{1}{2E_I\ms{(\bp)}}\,
\int\!\frac{[d^3\bP_F]}{[2E_F]}\delta^3(\bP_F-\bp)\,
|\cM_\alpha|^2
\Big|\sum_j
\frac{U_{\beta j}U^{\dag}_{j\alpha}}
{\Delta E_{\alpha j}-i\Lambda}
\Big|^2
}
does not depend on $\bp$.

Firstly, the dependence on $\bp$ of Eq.\,\eqref{gconst:def} lies on
$\delta^3(~)$,
$|\cM_\alpha|^2$ and $\Lambda=\Gamma/2\gamma_I$, where $\gamma_I=E_I/M_I$.
Then, apply a change of variables $\{P_F\}\to \{P'_F=\Lambda^{-1}_I P_F\}$ such
that the two sets of variables are related by a Lorentz boost $\Lambda_I$ defined by
$(E_I,\bp)=\Lambda_I (M_I,\bs{0})$.
We can write
\eq{
\label{gconst:1}
g(\bp)=
\frac{1}{2E_I\ms{(\bp)}}\,
\int\!\frac{[d^3\bP'_F]}{[2E'_F]}\frac{\delta^3(\bP'_F)}{\gamma_I}\,
|\cM_\alpha'|^2
\gamma^2_I
\Big|\sum_j
\frac{U_{\beta j}U^{\dag}_{j\alpha}}
{\Delta E'_{\alpha j}-i\Gamma/2}
\Big|^2
\,,
}
where the primed variables refer to the rest frame of $I$. We have also used the
relations $\delta^3(\bP'_F)=\gamma_I\ms{(\bp)}\delta^3(\bP_F-\bp)$ (notice
$\bp'=\bs{0}$) and $\Delta E'_{\alpha j}=\gamma_I\Delta E_{\alpha j}$.
Therefore,
\eq{
\label{gconst:2}
g(\bp)=
\frac{1}{2M_I}\,
\int\!\frac{[d^3\bP'_F]}{[2E'_F]}\delta^3(\bP'_F)\,
|\cM_\alpha'|^2
\Big|\sum_j
\frac{U_{\beta j}U^{\dag}_{j\alpha}}
{\Delta E'_{\alpha j}-i\Gamma/2}
\Big|^2
=g(\bs{0})
\,,
}
\textit{i.e.}, $g(\bp)$ can be calculated assuming the parent particle $I$ is at
rest.
We then obtain Eq.\,\eqref{FvioAB:4}.

\section{Some calculations}
\label{ap:calc}

We describe here some calculations necessary to get from Eq.\,\eqref{P:mu:2} to
Eqs.\,\eqref{P:mu:A:f} and \eqref{P:mu:B:f}.
The details to obtain Eqs.\,\eqref{P:mu:DeltaE:i1} and \eqref{P:mu:DeltaE:ii1} are
also shown.

For type A violation, Eq.\,\eqref{P:mu:A:f}, considering $\nu_\beta$ massless and
the two families approximation \eqref{Pvio:2f}, we can rewrite Eq.\,\eqref{P:mu:2} as
\eqarr{
\label{P:mu:3}
\mathcal{P}_{\to\bnu_\alpha\nu_\beta^\zero}=
\delta_{e\beta}|U_{\mu i}U^*_{\alpha i}|^2
\frac{8\tilde{G}^2}{2M_\mu(2\pi)^5}
\!
\int\! d^3\bk\,
\frac{(\Delta E_{\bnu_i})^2}{4\Lambda^3}
p_\alpha k_\beta 
\TL^{\alpha\beta}(Q,\varepsilon;M_e,0)
\,,
}
where $\Delta E_{\bnu_i}=E_{\bnu_i}(\bk)-E_{\bnu_1}(\bk)$,
$\TL^{\alpha\beta}$ is defined in Eq.\,\eqref{Tab:L:def},
$Q=(M_\mu-\bar{E}_{\bnu}\ms{(\bk)},-\bk)$,
$\bar{E}_{\bnu}\ms{(\bk)}\equiv\frac{1}{2}(E_{\bnu_i}\ms{(\bk)}
+E_{\bnu_1}\ms{(\bk)} )$
and $\varepsilon=\Delta E_{\bnu_i}$; $M_e$ is the electron mass.
We can compare Eq.\,\eqref{P:mu:3} to Eq.\,\eqref{G:mu:2}.
It is then possible to use the approximate expression \eqref{Tab:L:e} because 
the approximation is only inadequate for $|\bk|\approx M_\mu/2$ which corresponds to
the endpoint of the neutrino spectrum. Near the endpoint, however, the emission
probability is suppressed by the function $\rho_{i1}$.
After an angular integration, a change of variables and use of
Eqs.\,\eqref{dist:numu} we obtain Eq.\,\eqref{P:mu:A:f}.
The function $\rho_{ij}$ in Eq.\,\eqref{rho:1} is obtained after the
manipulation $\Delta E_{\bnu_i}=(m^2_i-m^2_1)/2\bar{E}_{\bnu}$.

For type B violation, Eq.\,\eqref{P:mu:B:f}, considering $\bnu_\alpha$ massless
and the two families approximation \eqref{Pvio:2f}, we can rewrite
Eq.\,\eqref{P:mu:2} as
\eqarr{
\label{P:mu:4}
\mathcal{P}_{\to\bnu_\alpha^\zero\nu_\beta}=
\delta_{\mu\alpha}
|U_{\beta j}U^*_{e j}|^2
\frac{8\tilde{G}^2}{2M_\mu(2\pi)^5}
\int\! d^3\bk'\,
\frac{(\Delta E_{\nu_j})^2}{4\Lambda^3}
(p\ponto k')
\TL^{\alpha}_{~\alpha}(Q,\varepsilon;M_e,0)
\,,
}
where $\Delta E_{\nu_j}=E_{\nu_j}(\bk')-E_{\nu_1}(\bk')$, 
$Q=(M_\mu-\bar{E}_{\nu}\ms{(\bk')},-\bk')$,
$\bar{E}_{\nu}\ms{(\bk')}\equiv
\frac{1}{2}(E_{\nu_j}\ms{(\bk')}+E_{\nu_1}\ms{(\bk')} )$
and $\varepsilon=\Delta E_{\nu_j}$.
Comparing Eq.\,\eqref{P:mu:4} to Eq.\,\eqref{G:mu:3}, 
after an angular integration, a change of variables and use of
Eqs.\,\eqref{dist:nue} and \eqref{Tab:L:e}, we obtain Eq.\,\eqref{P:mu:B:f}.

For Eqs.\,\eqref{P:mu:A:int} and \eqref{P:mu:B:int} it is necessary to use
\eqarr{
\label{int:x23}
\int_0^1 dx\frac{\omega^2 x^2}{\omega^2+x^2+a^2}&=&
\omega^2[1-\sqrt{a^2+\omega^2}\arctan(\frac{1}{\sqrt{a^2+\omega^2}})]
\cr
\int_0^1 dx\frac{\omega^2 x^3}{\omega^2+x^2+a^2}&=&
\tfrac{1}{2}\omega^2[1-(a^2+\omega^2)\ln(1+\frac{1}{a^2+\omega^2})]
\,.
}

To obtain Eq.\,\eqref{P:mu:DeltaE:i1} we firstly insert Eq.\,\eqref{Delta:psi}
into Eq.\,\eqref{P:mu:DeltaE:2} and use the two families approximation.
Then, perform the integrals in $\bk'$ and $\bq$ by using $T^{\alpha\beta}_\Lambda$
in Eq.\,\eqref{Tab:L0:def}.
We can assume any dependence of the integrand on $\bp$ can be approximated to the
central value $\bp\approx 0$ except for
\eq{
\label{sigma:x}
\int\!d^3{\bp}\,\Big|\hat{\bk}\ponto\frac{\partial\psi(\bp)}{\partial \bp}\Big|^2
=\ml{\frac{1}{2}}\sigma^2_x\,.
}
Equation \eqref{P:mu:DeltaE:2} is then identical to Eq.\,\eqref{G:mu:2} with 
$T^{\alpha\beta}_\Lambda$ \eqref{Tab:L0:def} instead of $T^{\alpha\beta}$
and the inclusion, in the integrand, of the term
\eq{
\ml{\frac{1}{2}}\sigma^2_x(\Delta \rk_i)^2=
\ml{\frac{1}{2}}\sigma^2_x(\sqrt{\rk^2_i+(m^2_i-m^2_1)}-\rk_i)^2\,.
}
The desired result is obtained after the approximations
$T^{\alpha\beta}_\Lambda\approx T^{\alpha\beta}$ and $\rk_i\approx \rk_1$ followed
by the angular integration in $\hat{\bk}$ and the integral in $\rk_1$ instead of
$E$.

Equation \eqref{P:mu:DeltaE:ii1} is obtained in a similar way. Firstly, insert  
Eq.\,\eqref{Delta:E} into Eq.\,\eqref{P:mu:DeltaE:2} and use the two families
approximation. 
Then, we perform the integrals in $\bk'$ and $\bq$ by using
$\TL^{\alpha\beta}$ in Eq.\,\eqref{Tab:L:def} and take the
central value $\bp\approx 0$ except for
\eq{
\int\!d^3{\bp}\,\Big|\frac{\hat{\bk}\ponto\bp}{E_\mu(\bp)} \psi(\bp)\Big|^2
=\ml{\frac{1}{2}}\frac{\sigma^2_p}{M^2_\mu}\,.
}
The additional term in the integrand is now
\eq{
\ml{\frac{1}{2}}\frac{\sigma^2_p}{M^2_\mu}\,
\frac{(\Delta k_i)^2}{2\Lambda^2}
=\Big(\frac{\sigma_pW_\nu}{M_\mu\Gamma}\Big)^2
\frac{1}{W^2_\nu}(\sqrt{\rk^2_i+(m^2_i-m^2_1)}-\rk_i)^2
\,,
}
where the factor $2\Lambda^2$ comes from the definition of $\TL^{\alpha\beta}$ when
compared to $T^{\alpha\beta}$. After the approximation $\TL^{\alpha\beta}\approx
T^{\alpha\beta}$ and the integration in $\rk_1$ we obtain the desired result.

\section{Two particle phase space}
\label{ap:Tab}

The contribution of any two-particle phase space to muon decay can be quantified
through
\eq{
\label{Tab:def}
T^{\alpha\beta}\ms{(Q;m_1,m_2)}
\equiv
\int\frac{d^3k_1}{2E_1}\frac{d^3k_2}{2E_2}
\delta^4\ms{(Q-k_1-k_2)}k_1^{\alpha}k_2^{\beta}
\,,
}
where $E_i=\sqrt{\bk_i^2+m^2_i}$, $k_i^0=E_i$, $i=1,2$. The integral in
Eq.\,\eqref{Tab:def} can be carried out straightforwardly by following some
steps: (1) rewrite
$\ml{\frac{d^3k_i}{2E_i}}=d^4k_i\,\delta(k^2_i-m^2_i)\theta(k_i^0)$, with free
$k_i^0$, and perform the change of variables $k_1=\lambda/2+x$, $k_2=\lambda/2-x$.
(2) Exploit the Lorentz covariance of the integral which ensures the property
$T^{\alpha\beta}\ms{(\Lambda^{-1} Q)}=
\ms{\Lambda^{\alpha}_{~\alpha'}\Lambda^{\beta}_{~\beta'}} T^{\alpha'\beta'}\ms{(Q)}$
for a general Lorentz transformation $\Lambda$. (3) From the property in (2), write
$T^{\alpha\beta}\ms{(Q)}=f_1\ms{(Q^2)}g^{\alpha\beta}+f_2\ms{(Q^2)Q^\alpha Q^\beta}$
and extract $f_1,f_2$ by calculating $T^{\alpha}_{~\alpha}$ and
$T^{\alpha\beta}Q_\alpha Q_\beta$.
The explicit form of $T^{\alpha\beta}$ is found to be
\eq{
\label{Tab:f}
T^{\alpha\beta}\ms{(Q;m_1,m_2)}
=
\frac{\pi}{24}\theta\ms{(Q^2\!-\!M^2)}A\,
\big[\ms{(g^{\alpha\beta}Q^2\!-\!Q^\alpha Q^\beta)}A^2
+\/ 3 \ms{Q^\alpha Q^\beta} B
\big]
\,,
}
where $M\equiv m_1+m_2$, $\Delta m\equiv m_1-m_2$ ~and
\eqarr{
\label{AB:def}\nonumber
A\ms{(Q^2)}&=&
\Big[1-\ml{\frac{M^2}{Q^2}}\Big]^{\meio}
\Big[1-\ml{\frac{(\Delta m)^2}{Q^2}}\Big]^{\meio}
\,, \\
B\ms{(Q^2)} &=&
\Big[1-\ml{\frac{M^2}{Q^2}}\ml{\frac{(\Delta m)^2}{Q^2}}\Big]
\,.
}
For massless particles, $T^{\alpha\beta}$ reduces to\,\cite{DGH}
\eq{
T^{\alpha\beta}\ms{(Q;0,0)}=
\frac{\pi}{24}\theta\ms{(Q^2)}\big(g^{\alpha\beta}Q^2-2Q^\alpha Q^\beta\big)
\,.
}

We can define a similar quantity for finite decay width, necessary in
Sec.\,\ref{ap:mudecay},
\eq{
\label{Tab:L:def}
\TL^{\alpha\beta}(Q,\varepsilon;m_1,m_2)
\equiv
\frac{2\Lambda^{\mss{3}}}{\pi}\!
\int\! {\frac{d^3\!k_1}{2E_1}\!\frac{d^3\!k_2}{2E_2}}
\frac{\delta^3\ms{(\bQ-\bk_1\!-\!\bk_2)}\,k_1^\alpha k_2^{\beta}}
{\ms{
|Q_0\!+\!\meio \varepsilon\!-\!E_1\!-\!E_2\!-\!i\Lambda|^2
|Q_0\!-\!\meio \varepsilon\!-\!E_1\!-\!E_2\!-\!i\Lambda|^2
}}
\,,
}
where $E_i=\sqrt{\bk_i^2+m^2_i}$, $Q=(Q_0,\bQ)$ is a four-vector, $\varepsilon$ is a
number and the factors in front of Eq.\,\eqref{Tab:L:def} are so chosen because
\eq{
\frac{2}{\pi}\frac{\Lambda^3}{[x^2+\Lambda^2]^2}
\rightarrow \delta(x)\,,
\text{ as $\Lambda\rightarrow 0$.}
}
In that way, 
\eq{
\TL^{\alpha\beta}\ms{(Q,0;m_1,m_2)}
\rightarrow
T^{\alpha\beta}\ms{(Q;m_1,m_2)}\,,
\text{ as $\Lambda\rightarrow 0$.}
}
The calculation of the integral \eqref{Tab:L:def} is more involved than
\eqref{Tab:def} because Lorentz covariance is lost but the step (1) described
previously can be carried out and leads to
\eq{
\label{Tab:L:1}
\frac{\pi}{2\Lambda^{\mss{3}}}
\TL^{\alpha\beta}
=
\frac{1}{4|\bQ|}
\int_{R}\!dx_0d\lambda_0d\varphi\,
\frac{
(\frac{\lambda}{2}+x)^\alpha
(\frac{\lambda}{2}-x)^\beta
}{
|Q_0\!+\!\meio\varepsilon\!-\!\lambda_0\!-\!i\Lambda|^2
|Q_0\!-\!\meio\varepsilon\!-\!\lambda_0\!-\!i\Lambda|^2
}
\,,
}
where $\lambda^\alpha=(\lambda_0,\bQ)$,
$\bx=x_{\parallel}\hat{\bQ}+\bx_{\perp}$, $\varphi$ is the angle in the
$\bx_\perp$ plane,
\eqarr{
x_{\parallel}&=&\bx\ponto\hat{\bQ}=\frac{1}{|\bQ|}(x_0\lambda_0-\mu^2)
\,,\\
|\bx_{\perp}|^2&=&\omega^2(\lambda_0,x_0,|\bQ|)
\,,\\
\omega^2(\lambda_0,x_0,|\bQ|)&\equiv&
-\frac{\lambda^2}{\bQ^2}(x_0-a)^2
+\ml{\frac{1}{4}}\lambda^2A^2(\lambda^2)
\,,\\
a&\equiv& \frac{\mu^2}{\lambda^2}\lambda_0\,.
}
The function $A(\lambda^2)$ is the same as \eqref{AB:def}.
The region of integration $R=R_1\cap R_2$ in the $(\lambda_0,x_0)$ plane is
constrained by
\eq{
\label{R:12}
R_1: ~\omega^2\ge 0,\quad R_2:~2|x_0|\le \lambda_0\,.
}
The constraint $R_2$ comes from the $\theta$-functions in step (1).
We can also consider the equivalent constraints coming from the original variables
\eq{
\label{l2:x2}
R_3:~\lambda^2=(k_1+k_2)^2\ge M^2\,, ~~\lambda_0\ge 0\,,
\quad
R_4:~(2x)^2=(k_1-k_2)^2\le (\Delta m)^2\,,
}
where $M,\Delta m$ are defined after Eq.\,\eqref{Tab:f}. The constraints
\eqref{l2:x2} follows from the fact that the variables $k_1,k_2$ are forward
time-like 4-vectors.

Analyzing the constraints in Eqs.\,\eqref{R:12} and \eqref{l2:x2} we conclude that
the region of integration involves
\eq{
\lambda_0\in [(\lambda_0)_{\rm min},\infty)\,,\qquad
x_0\in [(x_0)_{\rm min},(x_0)_{\rm max}]\,,
}
where
\eqarr{
(\lambda_0)_{\rm min}&=&\sqrt{M^2+\bQ^2}\equiv E_M(\bQ)\,, \\
(x_0)_{\underset{\mss{\rm min}}{\rm max}}
&=& a\pm\meios|\bQ|A(\lambda^2)\,.
}

It can be explicitly checked that the antisymmetric part is null
\eq{
\int_{\mt{(x_0)_{\rm min}}}^{\mt{(x_0)_{\rm max}}}
\hs{-2em}dx_0\!\int_0^{2\pi}\!d\varphi\,
\ml{\frac{1}{2}}(x^\alpha\lambda^\beta-\lambda^\alpha x^\beta)=0\,.
}

After some lengthy calculations we obtain
\eq{
\label{Tab:L:2}
\TL^{\alpha\beta}\ms{(Q,\mn{\varepsilon};m_1,m_2)}
=
\frac{\Lambda^3}{12}
\int_{\mt{(\lambda_0)_{\rm min	}}}^\infty\hs{-1.5em}d\lambda_0\,
A\ms{(\lambda^2)}
\frac{
[(g^{\alpha\beta}\lambda^2-\lambda^\alpha\lambda^\beta)A^2\ms{(\lambda^2)}
+3\lambda^\alpha\lambda^\beta B\ms{(\lambda^2)}]
}{
|Q_0\!+\!\meio\varepsilon\!-\!\lambda_0\!-\!i\Lambda|^2
|Q_0\!-\!\meio\varepsilon\!-\!\lambda_0\!-\!i\Lambda|^2
}
\,,
}
where $\lambda^2=\lambda^2_0-\bQ^2$ and $A,B$ are the same functions in
Eq.\,\eqref{AB:def}.
Since the integrand in Eq.\,\eqref{Tab:L:2} is only
appreciable when $\lambda_0\approx Q_0\gtrsim(\lambda_0)_{\rm min}$, provided
that $|\varepsilon|\ll\Lambda$, we can approximate 
\eq{
\label{Ap:d}
\text{Ap(d): } 
\left\{
\begin{array}{ll}
\lambda_0\approx Q_0\,, & \text{if } Q_0\ge E_M(\bQ),
\cr
\lambda_0= E_M(\bQ)\,, & \text{if } Q_0<E_M(\bQ),
\end{array}
\right.
}
in all the terms except in the denominator and notice 
$E_M(\bQ)=(\lambda_0)_{\rm min}$.
The remaining denominator can be integrated exactly and we obtain
\eq{
\label{Tab:L:f}
\TL^{\alpha\beta}\ms{(Q,\varepsilon;m_1,m_2)}
=
\frac{\Lambda^2}{\Lambda^2+\ms{\frac{1}{4}}\varepsilon^2}
H\ms{(\tfrac{1}{2}\varepsilon,\Lambda,\mfn{Q_0\!-\!E_M(\bQ)})}
\,T^{\alpha\beta}\ms{(Q;m_1,m_2)}
\,,
}
where $T^{\alpha\beta}$ is understood as the expression in
Eq.\,\eqref{Tab:f} without the $\theta$-function.
The role of the $\theta$-function is played by the function $H$ which is given by
\eqarr{
H(a,\Lambda,y)&\!=\!& \mL{\frac{1}{2}}
+\mL{\frac{1}{2\pi}}\arctan\!\Big(\!\ml{\frac{y-a}{\Lambda}}\!\Big)
+\mL{\frac{1}{2\pi}}\arctan\!\Big(\!\ml{\frac{y+a}{\Lambda}}\!\Big)
\cr &&\quad
-\ \mL{\frac{\Lambda}{4\pi
a}}\ln\!\Big[\ml{\frac{(y-a)^2+\Lambda^2}{(y+a)^2+\Lambda^2}}\Big]
\,.
}
Such function has the property that it is symmetric and localized around $0$ in
$a$-direction and it behaves as a smooth $\theta$-function (step function) in the
$y$-direction.
More specifically, $H(a,\Lambda,y)$ is negligible if $y\pm a\ll -\Lambda$,
$H(a,\Lambda,y)\sim 1-0$ if $y\pm a\gg \Lambda$, and
$H(a,\Lambda,y)\approx 1/2$ if $y^2-a^2\ll -\Lambda$.

Due to the properties of $H$, we can use the approximation
\eq{
\label{Ap:e}
\text{Ap(e): }\quad
\ml{\frac{\Lambda^2}{\Lambda^2+\ms{\frac{1}{4}}\varepsilon^2}}
H(\tfrac{1}{2}\varepsilon,\Lambda,\ms{Q_0-E_M(\bQ)})
\approx 
\ml{\frac{\Lambda^2}{\Lambda^2+\ms{\frac{1}{4}}\varepsilon^2}}
\theta(Q^2-M^2)\,,
} 
and obtain
\eq{
\label{Tab:L:e}
\TL^{\alpha\beta}\ms{(Q,\varepsilon;m_1,m_2)}
\approx
\frac{\Lambda^2}{\Lambda^2+\ms{\frac{1}{4}}\varepsilon^2}
\,T^{\alpha\beta}\ms{(Q;m_1,m_2)}
\,,
}
where now the $\theta$-function is included in $T^{\alpha\beta}$\,\eqref{Tab:f}.
The approximation Ap(e) in Eq.\,\eqref{Ap:e} is inadequate only around 
$\varepsilon\approx 0$ and $Q_0\approx E_M(\bQ)$ in a region of size
$\approx 2\Lambda$.
Notice Eq.\,\eqref{Tab:L:e} satisfies
$\TL^{\alpha\beta}\ms{(Q,0;m_1,m_2)}=T^{\alpha\beta}\ms{(Q;m_1,m_2)}$.

We can check the approximate expression in Eq.\,\eqref{Tab:L:e} is valid for 
$Q_0-E_M(\bQ)\gg \Lambda$ by noting that we can extend the lower
integration limit of Eq.\,\eqref{Tab:L:2} to $-\infty$ without changing the integral
appreciably. 

We made use of the integral
\eq{
\label{int:1}
\int_{-y}^{\infty}\!dx\,
\frac{1}
{[(x-a)^2+\Lambda^2]
[(x+a)^2+\Lambda^2]}
=
\frac{\pi}{2\Lambda}\frac{1}{a^2+\Lambda^2}
H(a,\Lambda,y)\,.
}
The integral can be performed by splitting the integrand into two terms, each
containing exclusively one of the factors of the denominator.
The integral in the limit $y\to\infty$ can be calculated explicitly by residues.

We see the tensorial form of $T^{\alpha\beta}_\Lambda$ \eqref{Tab:L:f} is
the same as $T^{\alpha\beta}$ \eqref{Tab:f} within the approximation Ap(d)
\eqref{Ap:d}.

For completeness, we also calculate the tensor 
\eq{
\label{Tab:L0:def}
\TT^{\alpha\beta}(Q;m_1,m_2)
\equiv
\frac{\Lambda}{\pi}\!
\int\! {\frac{d^3\!k_1}{2E_1}\!\frac{d^3\!k_2}{2E_2}}
\frac{\delta^3\ms{(\bQ-\bk_1\!-\!\bk_2)}\,k_1^\alpha k_2^{\beta}}
{\ms{
|Q_0\!-\!E_1\!-\!E_2\!-\!i\Lambda|^2
}}
\,,
}
where $E_i=\sqrt{\bk_i^2+m^2_i}$, $Q=(Q_0,\bQ)$ is a four-vector, and the factors in
front of Eq.\,\eqref{Tab:L0:def} are so chosen because of the limit \eqref{DeltaE=0}.
The detailed calculation is similar to the calculation of
$\TL^{\alpha\beta}(Q,\varepsilon;m_1,m_2)$ \eqref{Tab:L:def} above.
The final result within the approximation Ap(d) in Eq.\,\eqref{Ap:d} is
\eq{
\label{Tab:L0:1}
\TT^{\alpha\beta}(Q;m_1,m_2)\approx T^{\alpha\beta}(Q;m_1,m_2)
H_0(\ml{\frac{Q_0-E_M(\bQ)}{\Lambda}})\,,
}
where $T^{\alpha\beta}(Q;m_1,m_2)$ should be understood without the function
$\theta$ which is replaced by
\eq{
H_0(x)\equiv \ml{\frac{1}{2}}+
\ml{\frac{1}{\mn{\pi}}}\arctan(x)\,.
}

\section{Muon decay}
\label{ap:mudecay}

Let us consider the muon decay $\mu^+(p)\to \bnu_\mu(k) e^+(q)\nu_e(k')$, where the
4-momenta of each particle is explicitly written.

Muon decay is described by the four-point Fermi interaction
\eq{
\label{LF}
\lag=
-2\sqrt{2}G_F\sum_{\alpha,\beta,i,j}\Big(\bar{l}_\alpha(x)\gamma^\mu L U_{\alpha
i}\nu_i(x)\Big)
(\bnu_j(x)U^\dag_{j\beta}\gamma_\mu L l_\beta(x)\Big)\,,
}
where $L=\frac{1}{2}(1-\gamma_5)$ and $\{U_{\alpha i}\}$ denotes the PMNS matrix.
The invariant amplitude at tree level is given by
\eqarr{
\label{M:mu}
\cM_{ij}&=&2\sqrt{2}G_FU_{\mu i}U^*_{ej}\tilde{\cM}_{ij}\,,\\
\label{tM:mu}
\tilde{\cM}_{ij}&=&\bar{v}_\mu(\bp)\gamma^\alpha Lv_{\nu_i}(\bk)
\bar{u}_{\nu_j}(\bk')\gamma_\alpha Lv_{e}(\bq)\,,
}
while the square modulus, averaged over initial spin states and summed over final
spin states, is given by
\eq{
\mn{\frac{1}{2}}\sum_{\rm spins}
|\tilde{\cM}_{ij}|^2=
8(p\ponto k')(q\ponto k)\,.
}

The decay rate at rest is given by
\eqarr{
\label{G:mu:1}
\Gamma\ms{(\mu\!\rightarrow\!\bnu_\mu^\zero e\nu_e^\zero)}&=&
\frac{8\tilde{G}^2}{2M_\mu(2\pi)^5}\!
\int\!\frac{d^3\bq}{2E_e}
\frac{d^3\bk}{2E_{\bnu_\mu}}\frac{d^3\bk'}{2E_{\nu_e}}
\delta^4(p-P_F)\,
(p\ponto k')(q\ponto k)
\,,
\\
\label{G:mu:2}
&=&
\frac{8\tilde{G}^2}{2M_\mu(2\pi)^5}\!
\int\!
\frac{d^3\bk}{2E_{\bnu_\mu}}
p_\alpha k_\beta T^{\alpha\beta}(p-k;M_e,0)
\,,
\\
\label{G:mu:3}
&=&
\frac{8\tilde{G}^2}{2M_\mu(2\pi)^5}\!
\int\!
\frac{d^3\bk'}{2E_{\nu_e}}
(p\ponto k')T^{\alpha}_{~\alpha}(p-k';M_e,0)
\,,
\\
\label{Gmu}
&=&\frac{M^5_\mu G^2_F}{192\pi^2}\equiv \Gamma_\mu
\,,
}
where $p=(M_\mu,\bs{0})$ and $P_F=(q)_e+(k)_{\bnu_\mu}+(k')_{\nu_e}$.
The tensor $T^{\alpha\beta}$ is defined in Eq.\,\eqref{Tab:def}.

We can rewrite Eqs.\,\eqref{G:mu:2} and \eqref{G:mu:3} in terms of the energy
distributions of $\bnu_\mu$ and $\nu_e$, respectively, as
\eq{
\Gamma_\mu
\int_0^1 dx_{\bnu_\mu}
\cS_{\bnu_\mu}(x_{\bnu_\mu})
=
\Gamma_\mu
\int_0^1 dx_{\nue}
\cS_{\nue}(x_\nue)
\,.
}
The energy spectra for $e^+,\nu_e,\bnu_\mu$ at tree level, for $m_\nu=M_e=0$
are\,\cite{greub:plb93}:
\eqarr{
\label{dist:e}
\cS_e(x_e)&=&
2x^2_e(3-2x_e)\,,\\
\label{dist:numu}
\cS_{\bnu_\mu}(x_{\bnu_\mu})
&=&
2x^2_\bnumu(3-2x_\bnumu)\,,\\
\label{dist:nue}
\cS_\nue(x_\nue)
&=&
12 x^2_\nue(1-x_\nue)\,,
}
where $x_e\equiv E_e/W_e$, $x_\nue\equiv E_\nue/W_\nue$,
$x_\bnumu\equiv E_\bnumu/W_\bnumu$, $W_e\approx W_\nue\approx W_\bnumu\approx
M_\mu/2$; $W_{(~)}$ is the maximum energy.
The distributions \eqref{dist:e}--\eqref{dist:nue} can be obtained from
Eqs.\,\eqref{G:mu:1}--\eqref{G:mu:3}.


\section{Flavor diagonal term}
\label{ap:aa}

We intend here to analyze the second term in the r.h.s of
Eq.\,\eqref{sum23} which is flavor diagonal. We have estimated in
Ref.\,\onlinecite{intrinsic} that such term would be negligible for pion decay.
However, Eq.\,\eqref{P:tot:1} indicates that such term has to be of the order of
the flavor violating terms but negative in sign.

For concreteness, let us consider the muon decay $\mu^+\to\bnu_\alpha
e^+\nu_e^\zero$, neglecting the mass of one of the neutrinos.
Let us also disregard all the channels except the dominant one involving the
positron and neutrinos such that $\tilde{\mathcal{P}}=\mathcal{P}$ in
Eq.\,\eqref{P:tot:1}.
We then have for Eq.\,\eqref{P:tot:1},
\eq{
\mathcal{P}_{\to \mu\bnu_e}
+\mathcal{P}_{\to \mu\bnu_\mu}
+\mathcal{P}_{\to \mu\bnu_\tau}
=1\,.
}
Therefore,
\eq{
\label{P:mumu}
\mathcal{P}_{\to \mu\bnu_\mu}
=1
-\mathcal{P}_{\to \mu\bnu_e}
-\mathcal{P}_{\to \mu\bnu_\tau}
\,,
}
and the flavor conserving probability deviates from unity ($\mathcal{P}_{\rm tot}$)
by the flavor violating probabilities calculated in Sec.\,\ref{sec:mudecay}.

For comparison, let us rewrite the second term in the r.h.s of
Eq.\,\eqref{sum23} as
\eqarr{
\label{reFF}
2\re F^*_{\alpha 1}\Delta F_{\alpha i}&=&
\frac{2(\Delta E_{\bnu_i})\Delta E_{\alpha i}}
{|\Delta E_{\alpha i}-i\Lambda|^2|\Delta E_{\alpha 1}-i\Lambda|^2}
\cr
&=&
\frac{-(\Delta E_{\bnu_i})^2+2\Delta E_{\bnu_i}(E_\mu-\bar{E}_F)}
{|\Delta E_{\alpha i}-i\Lambda|^2|\Delta E_{\alpha 1}-i\Lambda|^2}
\,,
}
where $\Delta E_{\bnu_i}=E_{\bnu_i}-E_{\bnu_1}$ and
$\bar{E}_F=E_e+E_{\nu_e^\zero}+\bar{E}_{\bnu_i}$ is the mean final energy
considering $\bar{E}_{\bnu_i}=\frac{1}{2}(E_{\bnu_i}+E_{\bnu_1})$.
Wee see the first term in the numerator of Eq.\,\eqref{reFF} gives rise to the same
contribution, with opposite sign and different mixing matrix contribution, as
Eq.\,\eqref{Pvio:2f} considering Eq.\,\eqref{DeltaF:approx}.
On the other hand, the second term in the numerator of Eq.\,\eqref{reFF} was
estimated in Ref.\,\onlinecite{intrinsic} (appendix E) and it is of the order of
$\omega_{i1}^2(\Gamma/E_\nu)^2$, which is negligible. The same conclusion can be
reached by making the exact calculation considering Eq.\,\eqref{Tab:L:1} with an
additional $\lambda_0$ in the numerator.

For muon decay, within the two families approximation, we can calculate the
contributions from Eq.\,\eqref{sum23} explicitly,
\eqarr{
\label{P:mumu:1}
\mathcal{P}\ms{(\mu^+\!\to\!\bnu_\mu e^+\nu_e^{\mt{(0)}})}&=&
1-8|U_{\mu i}|^2\omega^2_{i1}+8|U_{\mu i}|^4\omega^2_{i1}
\\
&=&1-8|U_{\mu i}U_{\beta i}|^2\omega^2_{i1}\,,\quad \beta=e \text{ or } \tau\,.
}
The three terms of Eq.\,\eqref{P:mumu:1} correspond to the three terms
of Eq.\,\eqref{sum23}, in the same order, and we see Eq.\,\eqref{P:mumu} is
satisfied.



\begin{thebibliography}{99}
\bibitem{endnote0}
We are disregarding any consequences from anomalies.
\bibitem{endnote2}
The antineutrino states should be distinguished from the neutrino states in terms of
the helicity quantum number if neutrinos are of Majorana type.
\bibitem{endnote3}
Notice the genuine invariant amplitude is $U^*_{\alpha j}\cM_{\alpha j}$.
\bibitem{endnote1}
Although the formalism developed in Ref.\,\onlinecite{intrinsic} requires the
calculations to be performed assuming only momentum conservation,
but without energy conservation, we will explicitly assume the latter for the
calculation of the invariant amplitude since the possible deviation is quantified
by the decay width\,\cite{intrinsic} which is a tiny quantity
compared to other quantities in the problem.
\end{thebibliography}
\end{document}